\documentclass[usenatbib,useAMS]{mn2e}
\usepackage{times}
\usepackage{graphicx,mathptm}
\usepackage{longtable,afterpage}
\usepackage{subfigure}
\usepackage{url}
\usepackage{multirow}
\usepackage{amssymb}
\usepackage{lscape}
\usepackage[reqno]{amsmath}
\usepackage{color}
\usepackage[bookmarks=true,pdftitle={Classical Be stars in the Perseus Arm and beyond},pdfauthor={Roberto Raddi},pdfsubject={},pdfcreator={Roberto Raddi}, colorlinks=true, linkcolor=black, citecolor=black]{hyperref}
\usepackage[figure, figure*]{hypcap}

\def\aap{A\&A}

\def\aj{AJ}
\def\apj{ApJ}
\def\apjs{ApJS}

\def\mnras{MNRAS}
\def\pasp{PASP}
\def\araa{ARA\&A}

\defcitealias{SFD98}{SFD98}

\def\Halpha{{\rm H}\alpha}

\title[Classical Be stars in the Perseus Arm and beyond]{First results of an $\Halpha$ based search of classical Be stars in the Perseus Arm and beyond}
\author[Raddi et al.]{R. Raddi$^{1}$\thanks{E-mail: r.raddi1@herts.ac.uk}, J. E. Drew
$^{1}$, J. Fabregat$^{2}$, D. Steeghs$^{3}$,  N. J. Wright$^{4}$, S. E. Sale$^{5, 6, 7}$,
\newauthor  H. J. Farnhill$^{1}$, M. J. Barlow$^{8}$, R. Greimel$^{9}$, L. Sabin$^{10}$, R. M. L. Corradi$^{11,12}$, 
J. J. Drake$^{4}$\\
$^{1}$Centre for Astrophysics Research, STRI, University of Hertfordshire, College 
Lane Campus, Hatfield, AL10 9AB, U.K.\\
$^{2}$Observatorio Astron\'{o}mico, Universidad de Valencia, 46100 Burjassot, Spain\\
$^{3}$Department of Physics, University of Warwick, Coventry, CV4 9BU, U.K.\\ 
$^{4}$Smithsonian Astrophysical Observatory, 60 Garden Street, Cambridge, MA 02138, USA\\
$^{5}$ Departamento de F\'{i}sica y Astronom\'{i}a, Facultad de Ciencias, Universidad de Valpara\'{i}so, Av. Gran Breta\~na 1111, Playa Ancha, Casilla 53, Valpara\'{i}so, Chile\\
$^{6}$ Departamento de Astronom\'{i}a y Astrof\'{i}sica, Pontificia Universidad Cat\'olica de Chile, Av. Vicu\~na Mackenna 4860, Casilla 306,
Santiago 22, Chile\\
$^{7}$ Rudolf Peierls Centre for Theoretical Physics, Keble Road, Oxford OX1 3NP, UK\\
$^{8}$Department of Physics and Astronomy, University College London, Gower Street, London WC1E 6BT, UK\\
$^{9}$Institute for Geophysics, Astrophysics, and Meteorology, Institute of Physics, University of Graz,
Universitaetsplatz 5/II, 8010 Graz, Austria\\
$^{10}$Instituto de Astonom\'{i}a y Meteorolog\'{i}a, Departamento de F\'{i}sica, CUCEI, Universidad de Guadalajara, Av. Vallarta 2602, C.P. 44130, Guadalajara, Jal., Mexico\\
$^{11}$Instituto de Astrof\'{i}sica de Canarias, 38200 La Laguna, Tenerife, Spain\\
$^{12}$Departamento de Astrof\'{i}sica, Universidad de La Laguna, 38206 La Laguna, Tenerife, Spain\\}
\begin{document}

\date{Received 2012 June 12, Accepted 2013 January 7.}

\pagerange{\pageref{firstpage}--\pageref{lastpage}} \pubyear{}

\maketitle

\label{firstpage}

\begin{abstract}
We investigate a region of the Galactic plane, 
between $120^{\circ} \leq \ell \leq 140^{\circ}$ and 
$-1^{\circ} \leq b \leq +4^{\circ}$, and uncover a population of
moderately reddened $(E(B-V) \sim 1)$ classical Be stars
within and beyond the Perseus and Outer Arms. 370 candidate emission 
line stars $(13 \lesssim r \lesssim 16)$ selected from the INT Photometric 
$\Halpha$ Survey of the Northern Galactic plane (IPHAS) have been 
followed up spectroscopically. A subset of these, 67 stars with properties 
consistent with those of classical Be stars, have been 
observed at sufficient spectral resolution ($\Delta\lambda \approx 2$~--~4~\AA ) 
at blue wavelengths to narrow down 
their spectral types. We determine these to a precision estimated to be $\pm 1$
sub-type and then we measure reddenings via SED fitting with reference
to appropriate model atmospheres. Corrections for contribution to colour excess from
circumstellar discs are made using an established scaling to $\Halpha$
emission equivalent width. Spectroscopic parallaxes are obtained after
luminosity class has been constrained via estimates of distances to
neighbouring A/F stars with similar reddenings.  Overwhelmingly, the stars
in the sample are confirmed as luminous classical Be stars at
heliocentric distances ranging from 2~kpc up to $\sim$12~ kpc.  However, the
errors are presently too large to enable the cumulative distribution function 
with respect to distance to distinguish between models placing the stars 
exclusively in spiral arms, or in a smooth exponentially-declining 
distribution.
\end{abstract}

\begin{keywords}
stars: emission-line, early-type, Be - ISM: dust, extinction, structure - Galaxy: structure 
\end{keywords}

\section{Introduction}
\label{intro}

Outside the Solar Circle, the Perseus Arm is the first spiral arm
crossed by Galactic Plane sight-lines.  It contains a number
of well-studied star-forming clouds \citep[e.g. W3, W4 and W5,][]{SFR08}
set among stretches of relatively modest star-forming activity. 
The shape and characteristics of the Perseus Arm have been examined in 
several works over the years,  using different tracers ranging from 
CO \citep{Dame01} through to OB associations \citep{Russeil03}.  
A longstanding issue for these studies -- particularly in the second
quadrant of the Milky Way $(90^{\circ} \leq \ell \leq 180^{\circ})$ -- has been the evidence for peculiar
motions of stellar tracers and clouds, departing from the mean
rotation law, which necessarily challenge kinematic distance determinations
\citep[e.g.][]{Humphreys76, Carpenter00, Vallee08}.
Recently, distances to star forming regions within the Perseus and
other arms have begun to be measured reliably via methanol and OH maser 
trigonometric parallaxes, known to milli-arcsec precision.  Of special
significance to the present study is the \citet{Xu06} result for
W3OH ($\ell \simeq 134^{\circ}$) in the Perseus Arm, from which a
distance of 1.95 $\pm$ 0.04~kpc was obtained.  This represented a 
shortening of scale that has now been absorbed within the new
consensus as may be found in the works of \citet{Russeil07} and 
\citet{Vallee08}.

Beyond the Perseus Arm, also within the second quadrant, there is
some evidence accumulating in favour of the existence of a further spiral
arm, which is referred to as either the Outer or Cygnus Arm.
\citet{Russeil03}, \citet{Russeil07}, \citet{Levine06} and
\citet{SC10} have identified stellar and gaseous tracers,
that lend support to this outer structure. Nevertheless, its location
and true character remains elusive because of present
limits on the quantity of tracers available combined with the
continuing need to make significant use of kinematic distances.
For the Outer Arm, a distance between 5~-- 6~kpc, is quoted from fits of 
logarithmic spirals to the relevant tracers
\citep{Russeil03, Vallee08}. 
\citet{Negu03} also estimated a distance range running
from 5 to 6~kpc, via photometric parallaxes of a sample of bright OB 
stars.  The best single measurement to date is the maser parallax
obtained for WB89-437 by \citet{Hach09}, giving a distance of 
6.0$\pm$0.2~kpc.  At these distances, the Outer Arm straddles the zone of
Galacto-centric radii ($13 \lesssim R_G  \lesssim 14$~kpc) in which the
stellar disc 'truncates' \citep{Ruphy96} or, as has now become
clear, presents a pronounced shortening of exponential length scale 
\citep{Sale10}.

So whilst the reality of at least the Perseus Arm is beyond doubt, a 
settled picture of the Galactic Plane in the second quadrant is yet to
emerge.  In this paper, we add to the pool of available tracers a first 
sample of reddened classical Be (CBe) stars, reaching down to $r \approx
16$, that is drawn from the INT/WFC Photometric H$\alpha$ Survey of the Northern 
Galactic Plane (IPHAS)~\citep{Drew05} and in particular the catalogue 
of $\Halpha$ emission line sources provided in \citet{Witham08}.  In so doing we point towards 
the gain to be had from more comprehensive exploitation of these
newly available emission line objects.

CBe stars are mainly early B-type stars of luminosity class V-III that are 
on the Main-Sequence (MS) or moving off it \citep{Porter03}.  They are 
frequently observed in young open clusters ($\leq 30$~Myr)
\citep{Fabregat00}, and their spectra exhibit allowed transitions 
in emission (mainly lower excitation Balmer lines). Earlier CBe stars at least have not 
had time to move far from their birth places but, equally, they are 
unlikely to be heavily embedded in their parental clouds.
In addition they are intrinsically bright, with absolute magnitudes
ranging from $\sim 0$ down to $\sim -4$, enabling their detection at 
great distances across the Galactic Plane.  In combination, these 
attributes make them highly suitable targets for studying spiral arm 
structure.

We focus our study in a patch of sky, spanning $100\, \rm{deg}^{2}$, that
covers the Perseus Arm in the Galactic longitude range $120\degr \leq \ell
\leq 140\degr$ and latitude band $-1\degr \leq b \leq +4\degr$.
The positive offset of the chosen latitude band ensures we capture 
the displacement of the Galactic mid-plane caused by warping -- a 
phenomenon that is evident both from maps of H~I and dust emission 
\citep{Freudenreich94} and from the distribution of star forming 
complexes \citep{Russeil03}.  The selected longitude range encompasses 
the much-studied star forming complex W3/W4/W5, along with a more 
quiescent stretch of the Perseus Arm.  

We present the results of a two-stage spectroscopic follow-up 
programme.  The process begins with low resolution spectroscopy of 
$\sim$370 photometrically-selected candidate emission line objects -- 
the brighter portion of a total population in this part of the Plane, 
of more than 560 candidate emission line stars (Section~\ref{chap2}).  
This sample is further reduced to a set of 67 stars, for which we have
medium-resolution spectra that ultimately serve to confirm the
selected objects are nearly all luminous CBe stars.  In Section~\ref{chap3}, 
we determine spectral types and colour excesses for this sample and then 
estimate the contribution to the colour excess that originates in the circumstellar 
disc (that adds on to the interstellar component), which is observed toward each star.  
Using IPHAS survey data, we compare 
the resultant spectroscopic parallaxes with distances to similarly-reddened 
non-emission line A/F stars within a few arcminutes of each CBe star,
in order to set constraints on luminosity class.  This is described
in Section~\ref{chap4}, where we also present the spatial distribution 
of CBe stars that we obtain. Some of the sample appear to be very distant 
($R_{G} \geq 13$~kpc) early-type CBe stars.  The paper ends with a discussion 
that includes consideration of how the derived spatial distribution compares 
with simple simulations, accounting for typical errors, that place the 
stars either within the spiral arms only, or distributes them smoothly according
to an exponential stellar density profile.  We also consider
how the derived CBe star colour excesses compare with total integrated
values from the map of \citet[hereafter, \citetalias{SFD98}]{SFD98}.

\section{Spectroscopic follow-up of bright candidate emission line stars}
\label{chap2}
\subsection{Low resolution spectroscopy}
\label{chap2.1}
Candidate emission line stars in the specified Perseus Arm
region (Galactic longitude range $120\degr \leq \ell
\leq 140\degr$, latitude range $-1\degr \leq b \leq +4\degr$) were
identified from the \citet{Witham08} catalogue as potential spectroscopy
targets. All such objects are point sources that 
exhibit a clear $(r-\Halpha)$ excess, with respect to
main-sequence stars in the $(r-H\alpha,\, r-i)$ colour-colour diagram: 
560 such candidates fall within the chosen sky area, in the 
magnitude range $13 \leq r \leq 19.5$.  To enable
spectroscopic follow-up on small to mid-sized telescopes, we restricted
this sample to objects brighter than $r \approx 16$, i.e. 354 of them. 
To this list, we then added a further $\sim 50$ emission-line
candidates ($13 \lesssim r \lesssim 16$) derived 
from IPHAS photometry that was not available at the time 
the \citet{Witham08} catalogue was compiled.  

Observations of most of this moderately bright sample were collected 
between 2005 and 2011 at the 1.5m Fred Laurence Whipple Observatory (FLWO) 
Tillinghast Telescope using the FAST 
spectrograph \citep[][]{Fabricant98}. All in all, 370 objects were observed. The resolution of 
the spectra obtained was $\Delta\lambda \simeq 6$~\AA , and the data 
span the wavelength range 3500 -- 7500~\AA. The spectra from this 
facility were obtained in queue mode, and pipeline-processed  at the Telescope Data
Center at the Smithsonian Astrophysical Observatory. They were delivered without relative flux 
calibration.  An approximate calibration was applied to them subsequently, 
using a number of spectrophotometric standards taken from the FLWO-1.5m/FAST 
archive. 

Fig.~\ref{iphas} shows the IPHAS colours of the observed target stars that 
were confirmed by visual inspection of their spectra to be genuine $\Halpha$ emitters 
($> 90$\% of the 370 observed targets). The photometric colours are derived 
using an internal release of the forthcoming global calibration of IPHAS 
(Farnhill et al. in prep).
Fig.~\ref{map} shows the spatial distribution of the observed sample of 
targets. In both figures we pick out, in advance of discussion, the colours
and positions of the 67 CBe stars for which we have acquired mid-resolution 
spectra. 
\setcounter{figure}{0}
\begin{figure*}
\includegraphics[width=0.8\linewidth]{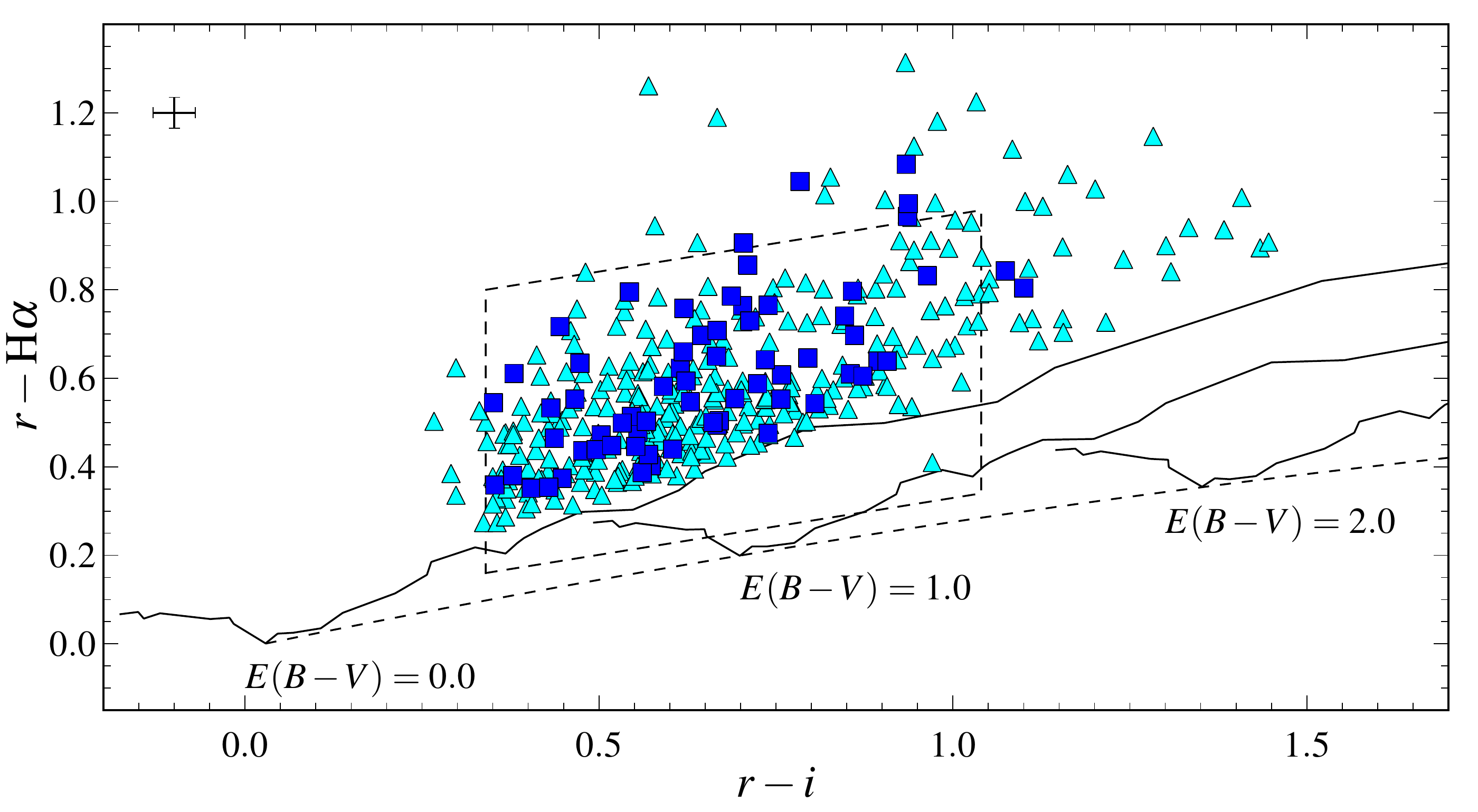}
\caption{IPHAS colour-colour diagram of the observed targets (cyan
triangles). Black solid lines are
  synthetic main sequence loci, at $E(B-V)$ = 0.0, 1.0, 2.0 
  \citep[see e.g. Table~2 in][]{Drew05}. These move parallel to the
  reddening vector that is plotted as the early-A reddening curve 
  (dashed lower curve). The box drawn above the
  unreddened main sequence defines the region in which CBe stars with
  $A_v \sim 4$ are likely to be located (cf Fig.~\ref{2mass} and the 
  discussion to be found in \citet{Corradi08}.The CBe stars, for which we have obtained intermediate-resolution
spectra, are picked out as blue squares. Typical error bars are plotted in the upper left corner. \label{iphas}}
\end{figure*}

\setcounter{figure}{1}
\begin{figure*}
\includegraphics[width=\linewidth]{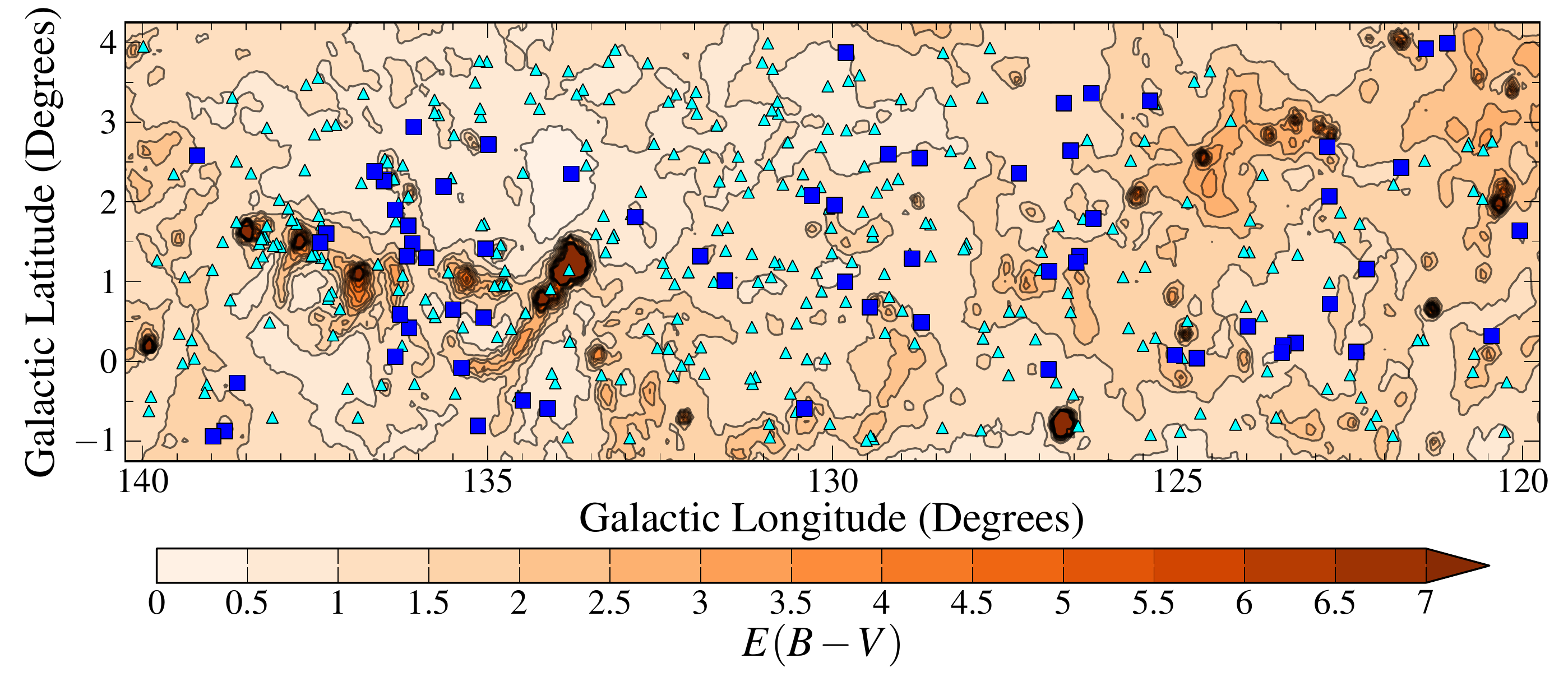}
\caption{A contour map of integrated dust column across the area, from 
\citetalias{SFD98}: the highest reddening contour drawn is for $E(B-V) = 7$. 
The spatial distribution of the candidate emission line stars is superposed. 
Symbols and colour scheme are the same as in Fig.~\ref{iphas}. \label{map}}
\end{figure*}

\subsection{Further reduction of the sample}
\label{chap2.2}
\setcounter{figure}{2}
\begin{figure}
\includegraphics[width=\linewidth]{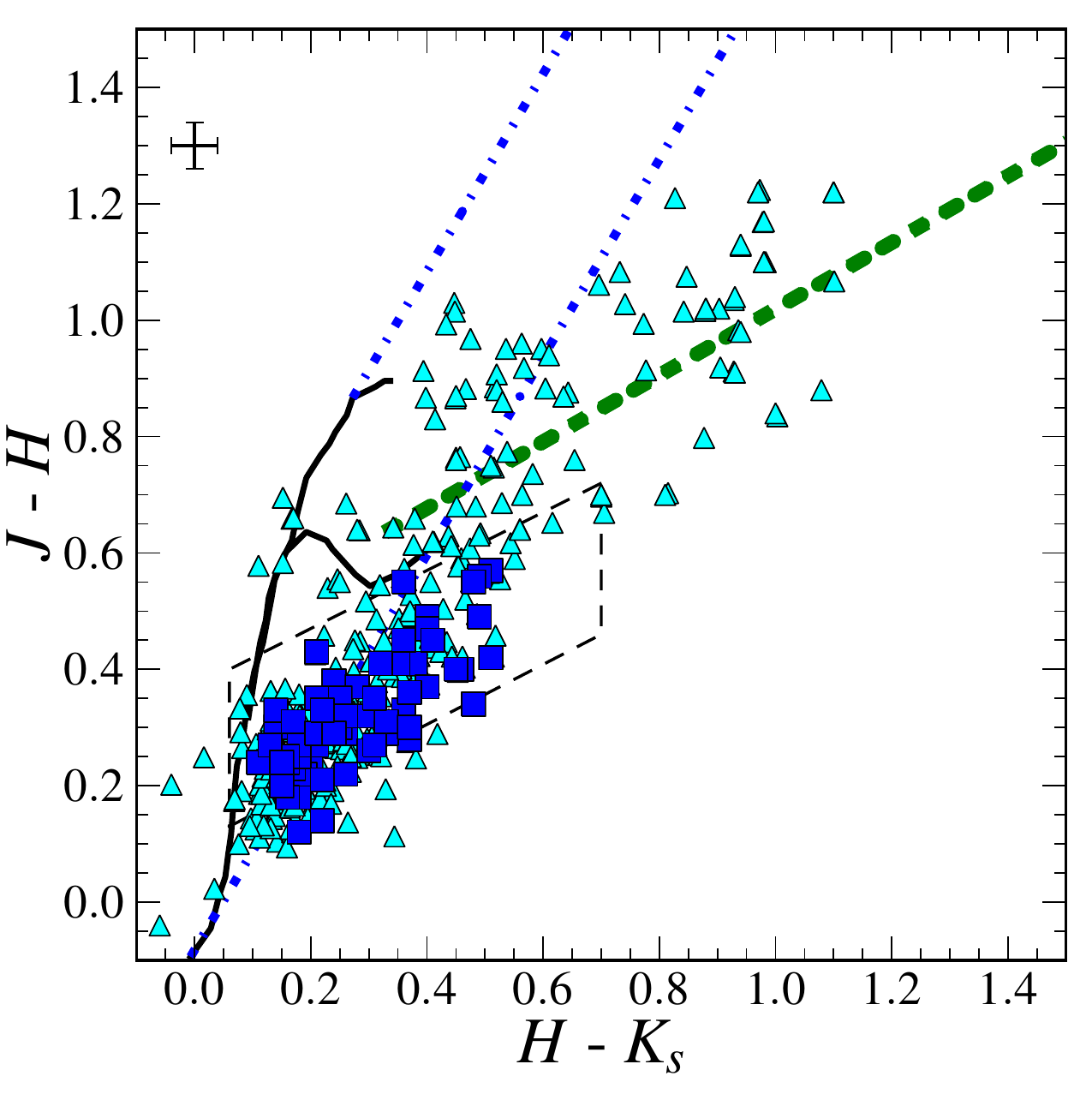}
\caption{2MASS near-infrared colour-colour diagram of our sample.  The colour
  symbols are as in Fig.~\ref{iphas}.  The solid line follows the
   dwarf and giant unreddened sequences \citep{Bessell88}. 
  The blue dot-dashed lines are the reddening vectors from 
  \citet{Rieke85}. The box drawn is due to \citet{Corradi08} and roughly
  delineates the region CBe stars with $A_V \sim 4$ would occupy, 
  while the green dashed line is the CTTS locus \citep{Meyer97}. 
  All the plotted curves are converted to the 2MASS system, adopting 
  relationships defined in \citet{Carpenter01}.\label{2mass}
  Typical error bars are in the upper left corner of the diagram.}
\end{figure}
The FLWO-1.5m/FAST observations have allowed us to cover
a large part of the potential target list for this region of sky, in order to
confirm/reject the emission line star status of the candidates, 
confirming a high success rate for IPHAS candidate emitter selection.  
The combination of achieved signal-to-noise ratio (S/N) and spectral 
resolution is sufficient for a first-pass coarse typing, allowing
early-type emission line stars to be clearly distinguished from
late-type.

The identification of suitable targets for further evaluation via 
intermediate
resolution spectroscopy, relied on two features that are frequently
shared with other classes of emission line star and so must be appraised 
carefully:

\begin{enumerate}

\item The bright $\Halpha$ emission, originating in the circumstellar
  environment of classical Be stars \citep{Porter03}, which is now
  known to originate from a disc \citep[][and references therein]{Dachs90}. 
  Similarly strong $H\alpha$ emission is also 
  observed both in low-mass YSOs, or classical T-Tauri stars (CTTS) 
  \citep{Bertout89}, and in intermediate-mass ones, or 
  Herbig~Ae/Be stars (HAeBe) \citep{Waters98}.  
  Very nearly all the confirmed emission line stars are either
  CBe stars or YSOs.  A property that can provide
  some discrimination is the presence/absence of nebular forbidden 
  line emission.  CBe-star spectra do not in general present with 
  forbidden line emission.  Any objects presenting such features are not
  included in the sample discussed here. 
\item A critical diagnostic separating CBe from candidate YSOs 
  is accessed at near-infrared (NIR) wavelengths. The spectral energy 
  distributions (SED) of optically-visible YSOs present a NIR colour
  excess due to thermal emission from a circumstellar disc.  The scale 
  of the excess depends on their evolutionary stage and type of object 
  \citep[see e.g.][]{Lada92, Meyer97}. But here the important
  point is that, by comparison with that of YSOs, the NIR excess 
  characteristic of CBe stars (due to circumstellar free-free
  emission) is very much weaker.

\end{enumerate}

We therefore supplemented IPHAS photometry and our low-resolution
spectroscopy with 2MASS photometry \citep{Cutri03} in order to help 
distinguish likely CBe stars from YSOs (and other emission line stars). We
required definite detections in all three 2MASS bands (quality flags
A, B or C) as minimum (i.e. 367 objects).  Then the measured $(J-H)$, 
$(H-K)$ colours must place the object in the domain close to or redward of
the line traced by non-emission early type stars as they redden.
There is some expectation that as $(H-K)$ grows, YSOs or similar
objects with strong NIR excesses may begin to mix in with the CBe
stars.    Our final sample of 67 CBe stars is drawn from those with
NIR colours bluer than $(J-H) = 0.6$. This removes
from consideration altogether, objects that may be red enough in
$(J-H)$ to be lightly-reddened CTTSs. 

Fig.~\ref{2mass} shows the 
2MASS colour-colour diagram of all potential targets: triangles 
distinguish objects with only low-resolution spectra and squares those 
deemed to be probable CBe stars that were selected for further 
spectroscopy at intermediate spectral resolution.  The number of
objects that could have satisfied all our selection criteria is
230 (out of 367).  The IPHAS and 2MASS photometry for
the sample of 67 objects scrutinised here is collected into 
Table~\ref{photometry}.

Our final selection exhibits the same broad range of reddenings,
$2 \lesssim A_{V} \lesssim 5$ present in the total available
sample: accordingly these objects' NIR colours are shifted parallel to 
the blue dot-dashed lines drawn in Fig.~\ref{2mass} that are themselves 
parallel to the reddening vector.  To underline this point we have
drawn in Fig.~\ref{2mass}, the $A_V \sim 4$ Be-star selection region
presented in the analysis of \citet{Corradi08} and note that
most of the target stars fall within it.  That the reddening is
significant is consistent with the presence of well-developed diffuse 
interstellar bands (DIB) in the spectra of the majority of our
selected targets.

Next, in Section~\ref{chap2.3}, we describe our
intermediate-resolution spectroscopy of this reduced sample and the 
reduction techniques that we adopted.
\setcounter{table}{0}
\begin{table*}
\begin{minipage}{150mm}
\caption[]{IPHAS and 2MASS photometry of the 67 CBe stars, with intermediate-resolution spectra. Objects will be identified by their
(\#) number, as given in this table in the rest of the paper. Columns are: 
IPHAS point-source name, which includes the J2000 RA and Dec; Galactic coordinates; $r$ magnitude, $(r - i)$ and $(r - \Halpha)$ colours from IPHAS;
 $J$ magnitudes, $(J-H)$ and $(H-K)$ colours from 2MASS. The errors on the $r$ magnitudes and the ($r-i$, $r-\Halpha$) colours are dominated
 by the photometric calibration and are respectively $0.02$, $0.03$, $0.035$. \label{photometry}}
\centering
\begin{tabular}{@{}rlrrrrrrrr@{}}
\hline
\#  & IPHAS & $\ell$ & {\it b} & $r$ & $(r-i)$& $(r - \Halpha)$   & $J$ & $(J - H)$& $(H - K)$\\
    & Jhhmmss.ss+ddmmss.s  &  (deg) & (deg)      &   (mag)   & (mag)  &       (mag)   & (mag)&(mag) & (mag) \\
\hline
 1 & J002441.73+642137.5 & 120.04 & 1.64  &    14.76  &  0.86  &  0.61 &$  12.51\pm0.02$ & $0.30\pm0.04$ & $0.26\pm0.04$\\
 2 & J002926.93+630450.2 & 120.45 & 0.32  &    14.07  &  0.35  &  0.36 &$  13.11\pm0.02$ & $0.12\pm0.04$ & $0.17\pm0.04$\\
 3 & J003248.02+664759.6 & 121.09 & 3.99  &    14.46  &  0.96  &  0.83 &$  12.11\pm0.02$ & $0.55\pm0.03$ & $0.36\pm0.03$\\
 4 & J003559.30+664502.9 & 121.40 & 3.92  &    15.96  &  0.76  &  0.61 &$  14.11\pm0.04$ & $0.40\pm0.06$ & $0.47\pm0.06$\\
 5 & J004014.89+651644.0 & 121.76 & 2.43  &    14.79  &  0.70  &  0.76 &$  12.95\pm0.02$ & $0.29\pm0.04$ & $0.26\pm0.05$\\
 6 & J004517.08+640124.1 & 122.26 & 1.16  &    15.62  &  0.89  &  0.64 &$  13.37\pm0.02$ & $0.41\pm0.04$ & $0.38\pm0.04$\\
 7 & J004651.69+625914.3 & 122.41 & 0.12  &    14.87  &  0.50  &  0.45 &$  13.32\pm0.02$ & $0.22\pm0.04$ & $0.26\pm0.05$\\
 8 & J005011.89+633525.8 & 122.79 & 0.72  &    15.37  &  0.62  &  0.62 &$  13.69\pm0.02$ & $0.28\pm0.04$ & $0.37\pm0.05$\\
 9 & J005012.69+645621.6 & 122.80 & 2.07  &    14.16  &  0.55  &  0.48 &$  12.65\pm0.03$ & $0.30\pm0.05$ & $0.14\pm0.05$\\
10 & J005029.25+653330.8 & 122.83 & 2.69  &    14.65  &  0.67  &  0.65 &$  12.97\pm0.03$ & $0.26\pm0.04$ & $0.30\pm0.04$\\
11 & J005436.84+630549.9 & 123.29 & 0.23  &    14.95  &  0.62  &  0.76 &$  13.30\pm0.03$ & $0.33\pm0.05$ & $0.36\pm0.05$\\
12 & J005611.62+630350.5 & 123.47 & 0.20  &    14.37  &  0.50  &  0.47 &$  12.95\pm0.02$ & $0.18\pm0.03$ & $0.18\pm0.03$\\
13 & J005619.50+625824.0 & 123.49 & 0.11  &    14.61  &  0.38  &  0.38 &$  13.46\pm0.02$ & $0.23\pm0.03$ & $0.20\pm0.04$\\
14 & J010045.58+631740.2 & 123.98 & 0.44  &    15.41  &  0.73  &  0.64 &$  13.53\pm0.02$ & $0.32\pm0.04$ & $0.29\pm0.05$\\
15 & J010707.68+625117.0 & 124.72 & 0.04  &    14.56  &  0.72  &  0.59 &$  12.65\pm0.02$ & $0.39\pm0.03$ & $0.24\pm0.03$\\
16 & J010958.80+625229.3 & 125.04 & 0.08  &    14.09  &  0.86  &  0.70 &$  12.44\pm0.02$ & $0.47\pm0.04$ & $0.40\pm0.04$\\
17 & J011543.94+660116.1 & 125.40 & 3.27  &    14.14  &  0.93  &  1.08 &$  11.95\pm0.02$ & $0.49\pm0.04$ & $0.40\pm0.04$\\
18 & J012158.74+642812.8 & 126.22 & 1.79  &    14.31  &  0.71  &  0.73 &$  12.74\pm0.03$ & $0.29\pm0.05$ & $0.33\pm0.05$\\
19 & J012405.42+660059.9 & 126.25 & 3.36  &    14.98  &  0.63  &  0.55 &$  13.45\pm0.03$ & $0.27\pm0.04$ & $0.21\pm0.04$\\
20 & J012320.10+635830.7 & 126.42 & 1.32  &    14.02  &  0.94  &  0.97 &$  11.89\pm0.02$ & $0.47\pm0.03$ & $0.40\pm0.03$\\
21 & J012339.76+635312.9 & 126.47 & 1.24  &    15.00  &  0.85  &  0.74 &$  12.96\pm0.02$ & $0.41\pm0.03$ & $0.35\pm0.03$\\
22 & J012609.27+651617.7 & 126.55 & 2.64  &    14.72  &  0.91  &  0.64 &$  12.79\pm0.03$ & $0.45\pm0.04$ & $0.35\pm0.04$\\
23 & J012751.29+655104.0 & 126.65 & 3.24  &    14.49  &  0.74  &  0.76 &$  12.77\pm0.02$ & $0.37\pm0.04$ & $0.28\pm0.04$\\
24 & J012703.24+634333.2 & 126.86 & 1.13  &    14.00  &  0.86  &  0.80 &$  11.60\pm0.02$ & $0.37\pm0.04$ & $0.40\pm0.04$\\
25 & J012540.54+623025.6 & 126.87 & -0.10 &    13.34  &  0.55  &  0.51 &$  12.05\pm0.02$ & $0.22\pm0.03$ & $0.19\pm0.02$\\
26 & J013245.66+645233.2 & 127.30 & 2.36  &    15.36  &  0.78  &  1.04 &$  13.31\pm0.03$ & $0.47\pm0.04$ & $0.40\pm0.05$\\
27 & J014218.74+624733.5 & 128.71 & 0.49  &    14.53  &  0.67  &  0.49 &$  12.90\pm0.03$ & $0.30\pm0.04$ & $0.23\pm0.04$\\
28 & J014620.44+644802.5 & 128.74 & 2.55  &    14.37  &  0.48  &  0.44 &$  13.25\pm0.03$ & $0.23\pm0.04$ & $0.17\pm0.04$\\
29 & J014458.14+633244.0 & 128.85 & 1.29  &    13.97  &  0.50  &  0.44 &$  12.89\pm0.02$ & $0.17\pm0.04$ & $0.16\pm0.04$\\
30 & J015037.67+644446.9 & 129.19 & 2.60  &    14.59  &  0.45  &  0.37 &$  13.52\pm0.03$ & $0.21\pm0.04$ & $0.23\pm0.04$\\
31 & J014905.18+624912.3 & 129.46 & 0.68  &    13.71  &  0.87  &  0.61 &$  11.49\pm0.02$ & $0.34\pm0.04$ & $0.48\pm0.04$\\
32 & J015918.32+654955.8 & 129.81 & 3.87  &    15.14  &  0.53  &  0.50 &$  13.80\pm0.02$ & $0.24\pm0.03$ & $0.16\pm0.04$\\
33 & J015246.27+630315.0 & 129.82 & 1.00  &    14.35  &  0.57  &  0.40 &$  12.90\pm0.02$ & $0.25\pm0.04$ & $0.18\pm0.04$\\
34 & J015613.22+635623.8 & 129.97 & 1.96  &    14.05  &  0.47  &  0.63 &$  12.82\pm0.03$ & $0.31\pm0.04$ & $0.33\pm0.04$\\
35 & J015922.53+635829.3 & 130.30 & 2.08  &    15.06  &  0.54  &  0.79 &$  13.37\pm0.02$ & $0.45\pm0.03$ & $0.36\pm0.03$\\
36 & J015427.15+612204.7 & 130.41 & -0.59 &    14.29  &  1.07  &  0.84 &$  11.55\pm0.02$ & $0.57\pm0.03$ & $0.51\pm0.02$\\
37 & J020734.24+623601.1 & 131.56 & 1.01  &    14.42  &  0.60  &  0.44 &$  12.98\pm0.02$ & $0.27\pm0.04$ & $0.18\pm0.04$\\
38 & J021121.67+624707.5 & 131.92 & 1.32  &    15.54  &  0.67  &  0.50 &$  13.93\pm0.03$ & $0.35\pm0.05$ & $0.25\pm0.05$\\
39 & J022033.45+625717.4 & 132.86 & 1.81  &    15.75  &  0.71  &  0.86 &$  13.90\pm0.02$ & $0.40\pm0.04$ & $0.45\pm0.04$\\
40 & J022953.82+630742.3 & 133.79 & 2.35  &    14.31  &  0.66  &  0.50 &$  12.85\pm0.02$ & $0.22\pm0.03$ & $0.15\pm0.03$\\
41 & J022337.05+601602.8 & 134.13 & -0.59 &    14.00  &  0.57  &  0.50 &$  12.54\pm0.02$ & $0.25\pm0.02$ & $0.16\pm0.03$\\
42 & J022635.99+601401.8 & 134.49 & -0.49 &    14.54  &  0.94  &  0.99 &$  12.13\pm0.02$ & $0.56\pm0.02$ & $0.49\pm0.03$\\
43 & J024054.96+630009.7 & 134.99 & 2.72  &    15.72  &  0.43  &  0.53 &$  14.36\pm0.03$ & $0.29\pm0.05$ & $0.21\pm0.07$\\
44 & J023642.66+614714.9 & 135.03 & 1.41  &    15.44  &  0.56  &  0.39 &$  13.95\pm0.05$ & $0.33\pm0.07$ & $0.14\pm0.06$\\
45 & J023404.70+605914.4 & 135.06 & 0.55  &    12.91  &  0.62  &  0.66 &$  11.30\pm0.02$ & $0.42\pm0.03$ & $0.52\pm0.03$\\
46 & J023031.39+594127.1 & 135.14 & -0.81 &    14.49  &  0.57  &  0.43 &$  12.96\pm0.02$ & $0.31\pm0.04$ & $0.17\pm0.04$\\
47 & J023431.07+601616.6 & 135.38 & -0.08 &    13.62  &  1.10  &  0.80 &$  10.77\pm0.02$ & $0.55\pm0.03$ & $0.48\pm0.04$\\
48 & J023744.52+605352.8 & 135.50 & 0.65  &    16.79  &  0.74  &  0.48 &$  14.76\pm0.04$ & $0.28\pm0.07$ & $0.31\pm0.09$\\
49 & J024405.38+621448.7 & 135.64 & 2.19  &    15.24  &  0.44  &  0.46 &$  13.80\pm0.02$ & $0.49\pm0.03$ & $0.49\pm0.04$\\
50 & J024252.57+611953.9 & 135.89 & 1.30  &    15.75  &  0.70  &  0.91 &$  13.95\pm0.03$ & $0.36\pm0.05$ & $0.37\pm0.05$\\
51 & J025016.66+624435.6 & 136.07 & 2.94  &    14.53  &  0.40  &  0.35 &$  13.21\pm0.02$ & $0.24\pm0.04$ & $0.11\pm0.04$\\
52 & J024504.86+612502.0 & 136.09 & 1.48  &    15.38  &  0.59  &  0.58 &$  13.81\pm0.02$ & $0.35\pm0.04$ & $0.21\pm0.04$\\
53 & J024146.74+602532.2 & 136.14 & 0.42  &    14.06  &  0.64  &  0.70 &$  12.31\pm0.02$ & $0.30\pm0.03$ & $0.25\pm0.03$\\
54 & J024618.12+613514.7 & 136.15 & 1.70  &    15.60  &  0.47  &  0.55 &$  14.33\pm0.03$ & $0.27\pm0.04$ & $0.13\pm0.05$\\
55 & J024506.09+611409.1 & 136.17 & 1.32  &    15.97  &  0.69  &  0.79 &$  14.12\pm0.02$ & $0.35\pm0.03$ & $0.31\pm0.04$\\
56 & J024317.68+603205.5 & 136.27 & 0.59  &    13.69  &  0.67  &  0.71 &$  11.98\pm0.03$ & $0.32\pm0.04$ & $0.26\pm0.04$\\
57 & J024159.21+600106.0 & 136.34 & 0.06  &    14.56  &  0.69  &  0.55 &$  12.80\pm0.02$ & $0.29\pm0.03$ & $0.21\pm0.03$\\
\hline
\end{tabular}
\end{minipage}
\end{table*}
\begin{table*}
\begin{minipage}{150mm}
\contcaption{}
\centering
\begin{tabular}{@{}rlrrrrrrrr@{}}
\hline
\#  & IPHAS & $\ell$  &{\it b} & $r$ & $(r-i)$& $(r - \Halpha)$   & $J$ & $(J - H)$& $(H - K)$\\
    & Jhhmmss.ss+ddmmss.s  &  (deg) & (deg)      &   (mag)   & (mag)  &       (mag)   & (mag)&(mag) & (mag) \\
\hline
58 & J024823.69+614107.1 & 136.34 & 1.90  & 13.92  &  0.35  &  0.54 & $12.75\pm0.02$ & $0.29\pm0.03$ & $0.24\pm0.03$\\
59 & J025102.22+615733.8 & 136.50 & 2.28  & 14.10  &  0.45  &  0.72 & $12.70\pm0.02$ & $0.30\pm0.04$ & $0.37\pm0.04$\\
60 & J025059.14+615648.7 & 136.50 & 2.26  & 15.32  &  0.38  &  0.61 & $14.16\pm0.03$ & $0.21\pm0.04$ & $0.22\pm0.05$\\
61 & J025233.25+615902.2 & 136.64 & 2.38  & 14.82  &  0.43  &  0.35 & $13.60\pm0.03$ & $0.14\pm0.05$ & $0.22\pm0.06$\\
62 & J025448.85+605832.1 & 137.34 & 1.60  & 16.13  &  0.76  &  0.55 & $13.90\pm0.03$ & $0.41\pm0.05$ & $0.32\pm0.05$\\
63 & J025502.38+605001.9 & 137.43 & 1.49  & 14.48  &  0.52  &  0.45 & $12.96\pm0.02$ & $0.33\pm0.04$ & $0.22\pm0.04$\\
64 & J025704.89+584311.7 & 138.63 & -0.27 & 16.22  &  0.80  &  0.54 & $14.13\pm0.02$ & $0.43\pm0.03$ & $0.21\pm0.04$\\
65 & J025610.40+580629.6 & 138.81 & -0.87 & 13.79  &  0.55  &  0.45 & $12.30\pm0.02$ & $0.20\pm0.03$ & $0.15\pm0.03$\\
66 & J025700.49+575742.8 & 138.98 & -0.94 & 14.26  &  0.62  &  0.59 & $12.60\pm0.02$ & $0.24\pm0.03$ & $0.15\pm0.03$\\
67 & J031208.92+605534.5 & 139.21 & 2.58  & 15.12  &  0.79  &  0.65 & $12.90\pm0.02$ & $0.45\pm0.03$ & $0.41\pm0.03$\\

\hline
\end{tabular}
\end{minipage}
\end{table*}
\setcounter{table}{1}
\begin{table*}
\begin{minipage}{160mm}
\caption{La Palma observations and relevant telescope set-up information,
  sorted by date of observation. \label{logs}}
\begin{tabular}{@{}ccccccc@{}}
\hline
  Run & Telescope/Instrument & Grating &  Wavelength interval & 
$\Delta\, \lambda$ & Observed targets & Apparent magnitude ($r$) \\
\hline
2006-08-27/29, 2006-09-08 & INT/IDS & R300V & 3500-7500 $\rm{\AA}$ & $4 
\rm{\AA}$ & 32 & $\sim 14.0$~--~$16.0$ \\
2007-12-04/07 & NOT/ALFOSC & \#16 & 3500-5000 $\rm{\AA}$& $ 2 \rm{\AA}$ & 
26 & $\sim 13.5$~--~$17.0$ \\
2009-11-27/30   & INT/IDS & R400V & 3500-7500 $\rm{\AA}$ & $3 \rm{\AA}$ & 
2 & $\sim 13.0$~--~$14.0$ \\
2010-10-21/26 & INT/IDS & R400V & 3500-7500 $\rm{\AA}$ & $3 \rm{\AA}$ 
& 7 & $\sim 13.0$~--~$16.0$\\
\hline
\end{tabular}
\end{minipage}
\end{table*}
\subsection{La Palma Observations}
\label{chap2.3}
We obtained mid-resolution and high S/N spectra of the 67 selected 
targets, on La Palma at the Isaac Newton Telescope (INT), using the
Intermediate Dispersion Spectrograph (IDS), and on the
Nordic Optical Telescope (NOT) using the Andalucia Faint Object
Spectrograph and Camera (ALFOSC).  The data were obtained over 18
nights between the years of 2006 and 2010.  A further practical
criterion that came into play in deciding which of the probable CBe 
stars to prioritise for mid-resolution spectroscopy was to prefer 
objects for which $(B-r) \lesssim 2$ was anticipated, giving a better
prospect of a blue spectrum of usable quality.  As will become
apparent, this limited reddenings to $A_V \lesssim 5$, or equivalently
$E(B-V) \lesssim 1.7$. 

Relevant information about spectrograph set-ups for these
observations are listed in Table~\ref{logs}.  The main point of
contrast between the INT and NOT data is that a bluer, higher
resolution grating was chosen for the latter, offering  
better opportunities for traditional blue-range spectral-typing -- at
the price of no coverage of the H$\alpha$ region.  

To break this down a little further, three runs took place at the INT 
(semester B, 2006, 2009 and 2010), observing respectively 32, 2, and
7 objects with the IDS.  In 2006, we used the R300V grating, with a 
dispersion of $1.87\,\rm{\AA}/\rm{pix}$, while in the other two runs
we preferred R400V, giving $1.41\,\rm{\AA}/\rm{pix}$.  
During each run, the slit width was 1'' so as to achieve spectral resolutions
of, respectively,  
$\Delta\lambda \approx 4\, \rm{\AA}$ and $\Delta \lambda\approx 3 \rm{\AA}$. 
Both set-ups cover the blue-visible interval and extend into the far red, 
but the disturbance due to fringing at wavelengths longer than 
$\sim \lambda7500\, \rm{\AA}$ was sufficiently severe that in practice
we did not use the spectrum at these longer wavelengths.

Twenty--six spectra were observed with NOT/ALFOSC, in December 2007, using 
grating \#16, which gives a dispersion of $0.77\, \rm{\AA}/\rm{pix}$. 
The slit width was set to 0.45'',  in order to achieve a resolution of 
$\Delta \lambda \approx 2\, \rm{\AA}$. The wavelength interval covers
the blue spectrum, from the Balmer jump up to $\rm{H}\beta$. 

Data reduction - i.e. the standard steps of bias subtraction,
flat-fielding, sky subtraction, wavelength calibration, extraction 
and flux calibration - was accomplished by using standard IRAF
routines.

Spectrophotometric standards were observed across all the nights, with
a wider slit, to allow a relative flux calibration to be applied.
Also to enable this, all target stars were observed with the slit
angle set at the parallactic value.
An unfortunate choice of standards in the first INT run
prevented the construction of a validated flux calibration curve at 
wavelengths redder than $\lambda 5000$\,\AA . However, at shorter
wavelengths the several standard star observations available could be
combined to produce a well-validated correction curve.  For this
reason, and because it matches the wavelength range offered by the NOT
spectra, all spectrophotometric reddening estimates
(Section~\ref{chap3.1}) are based on fits to the spectrum shortward of 5000~\AA. 
  
Negligibly reddened spectral type standards were also observed from
time to time, and these provided us with some useful checks on the
final flux calibration applied to our data.  Based on these we
determine that the flux calibration itself will not introduce
reddening errors larger than $\Delta E(B-V) = 0.05$.
On most nights, arc lamps were acquired before and after each star was 
observed, and were subsequently used as the basis for wavelength 
calibration.  The wavelength precision achieved ranges between 
0.10 and 0.15$\rm{\AA}$.  

At least two exposures were obtained for each target in
order to mitigate ill effects from unfortunately-placed cosmic rays
but in many instances three or four exposures were collected to
improve the signal-to-noise ratio.  Individual exposure times ranged
from 300~sec for the brightest targets, up to 1500/1800~sec for the 
faintest. The S/N ratio, at 4500~\AA, ranges from 22 up to just over 
100, the median of the distribution being 45.

\section{Analysis of the intermediate resolution spectra}
\label{chap3}
From here on, the discussion focuses exclusively on the 67 probable 
CBe stars with mid-resolution ($1200 < R < 2400$) spectra.  First we
describe the classification of the spectra, and then present our two
methods for reddening determination.

\subsection{Spectral classification}
\label{chap3.1}
\setcounter{figure}{3}
\begin{figure}
\centering
\subfigure[]{\includegraphics[width=\linewidth]{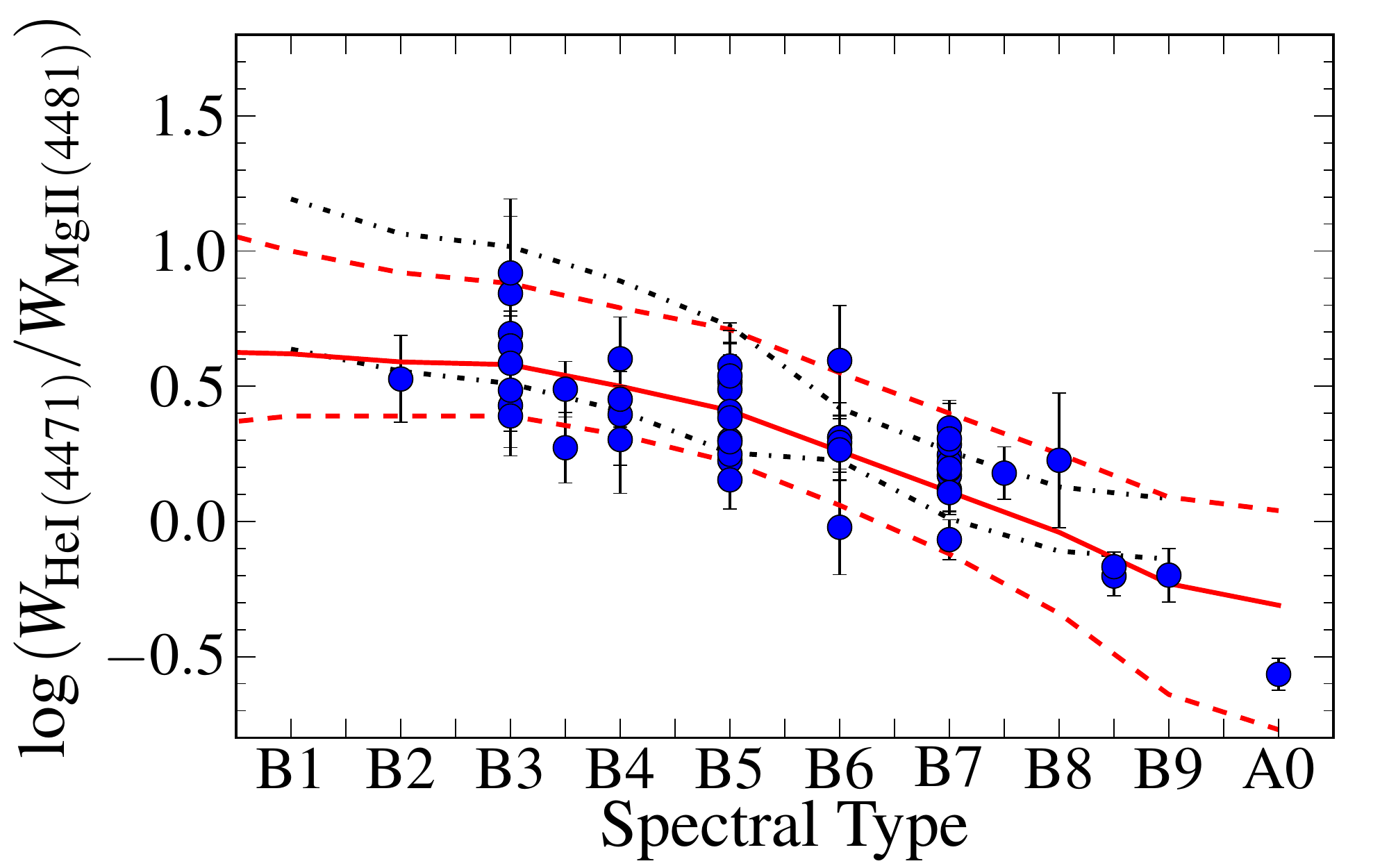}\label{spt:a}}
\subfigure[]{\includegraphics[width=\linewidth]{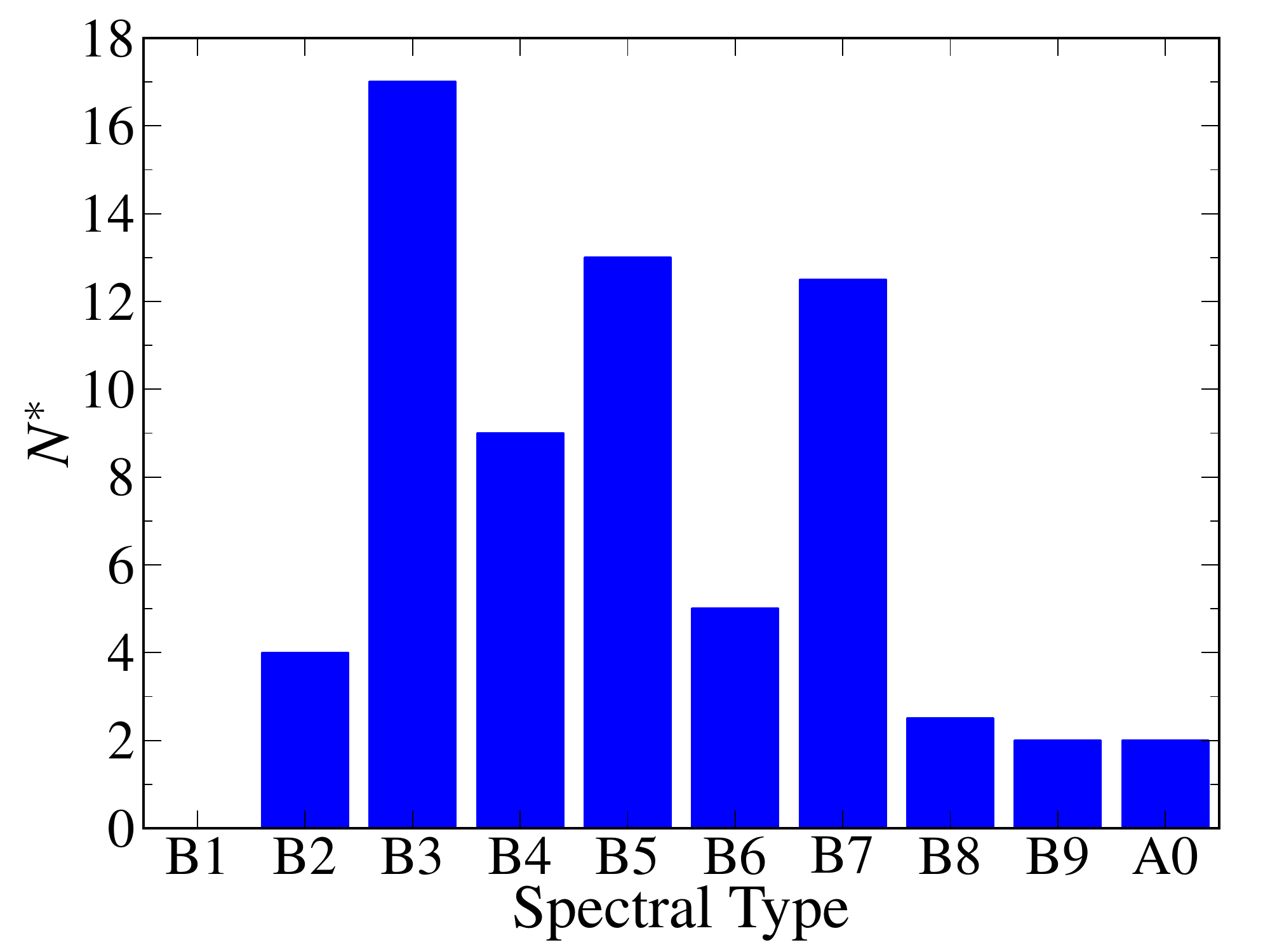}\label{spt:b}}
\caption{(a): Correlation between spectral type and the logarithm of the equivalent width ratio,  W(He~I $\lambda 4471$/Mg~II $\lambda 4481$. 
Our sample is compared to an interpolation of the high-resolution data from 
\citet{Chauv01}:  the black dotted curves mark the 1$\sigma$ confidence limits. Dashed red curves, instead, 
represent the 1$\sigma$ confidence limits obtained from repeated measures of 
equivalent-width ratios from appropriately chosen model atmospheres, with random noise added that matches S/N = 40.
The blue circles are the values obtained for our sample. 
(b): The histogram of spectral sub-types in the CBe sample.}
\label{spt}
\end{figure}
\setcounter{figure}{4}
\begin{figure}
\centering
\includegraphics[width=\linewidth]{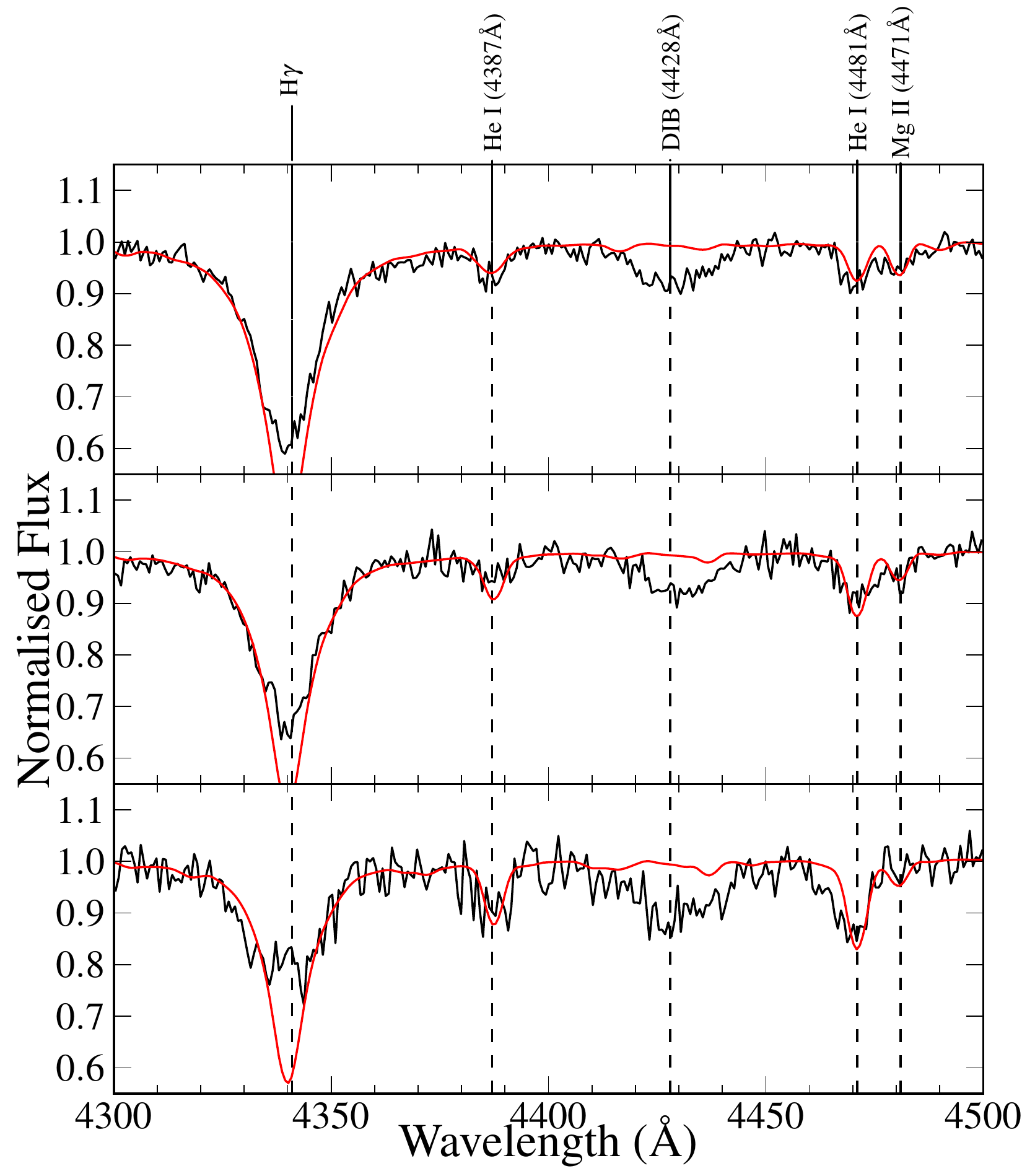}
\caption{Examples of spectral type assignments based on three spectra
with different $S/N$ ratio. From top to bottom,
\#41 (B7, $S/N = 64)$, \#65 (B5, $S/N =48)$, \#50 (B3, $S/N = 34)$. The 
observed spectra are in black while the preferred \citet{Munari05}
models are in red. The models have been rebinned to match that of the 
observations.}
\label{sn_fitting}
\end{figure}
Where $\Halpha$ was present in the wavelength range observed, it was
always seen in emission.  In the higher resolution NOT data, missing
the red part of the spectrum, we generally found the $\rm{H}\beta$
line to be either inverted or partially filled in.  This means that 
caution must be exercised in allowing the Balmer line profiles to
inform the classification of a star's spectrum. 

Spectral types were first determined, by direct comparison both with
spectral-type standards that we acquired during each observing run and also
with templates taken from the INDO-US library \citep{Valdes04}. The
latter needed to be degraded in spectral resolution from the original 
$\Delta \lambda = 1\, \rm{\AA}$ to match that of our data. 
Nearly all stars in the reduced sample of 67 were B stars exhibiting
He~I absorption, with only one or two crossing the boundary to A-type.  
No star showed He~II, ruling out any as O-type.  Our assignments were
guided by the criteria to be found in \citet{Jaschek87}, \citet{Gray09} and 
\citet{Didelon82}. The last of these usefully supplies
quantitative measures of equivalent width variation with spectral type and 
luminosity class.  Our list of key absorption lines for spectral type 
determination is:
\begin{itemize}
\item B-type: He~I lines at $\lambda \lambda 4009$-$4026\, \rm{\AA}$, $\lambda
\lambda 4121$-$4144\, \rm{\AA}$ and $\lambda \lambda\, 4387$-$4471 \rm{\AA}$ 
compared to the Mg~II $\lambda 4481\, \rm{\AA}$; 
\item A-type: Ca~II K and Mg~II.  The absence of He~I.
\end{itemize}
How well fainter features can be detected depends on the specifics of
the achieved S/N ratio and the spectral resolution -- and the first of
these depends in turn on how much interstellar extinction is present.
Because the reddening is significant, it is generally the case that
our classifications of the B stars depend heavily on the relative 
strengths of the He~I $\lambda 4471$ and Mg~II $\lambda 4481$ features --
a good $T_{\rm{eff}}$ indicator, with little sensitivity to $\log{g}$
within classes V-III -- rather than on shorter wavelength lines. As
young, thin-disc objects, CBe stars are unlikely to present with
distinctive blue spectra indicating significant metallicity variation,
even to quite large heliocentric distances. So we make no attempt at 
this stage to treat metallicity as a detectable variable.
 
Furthermore, in CBe stars, the above mentioned transitions can
be affected to differing extents by infilling line emission or 
continuum veiling due to the presence of ionised circumstellar discs,
while in faster rotators, line blending can also be an issue.  These
factors raise challenges to typing methods dependent on main 
sequence (MS) templates.  To overcome these problems, line 
equivalent widths ratios should also be brought into consideration, 
as these suffer less modification. 

As a way of refining our spectral typing, where possible, we measured
the absorption equivalent-width ratio $W_{\lambda 4471} / W_{\lambda 4481}$, via
simple gaussian fitting with the STARLINK/DIPSO tool, and compared it 
with data from \citet{Chauv01}  and model atmospheres, in Fig.~\ref{spt:a}. 
The model atmosphere predictions include simulated noise, corresponding to S/N = 40.
The precision of the typing, as judged by eye, is to $\pm$ one sub-type for all but the
lowest quartile in S/N ratio (S/N $<$ 35) where it approaches $\pm$2
sub-types (these objects have generally larger uncertainties,
$\Delta(\log{W_{\lambda 4471} / W_{\lambda 4481}}) \gtrsim 0.30$, and
are not plotted in Fig.~\ref{spt:a}).   Noisy spectra are subject to a
dual bias, depending on the actual value of the line ratio. Early-B
types, when the Mg~II line is weaker compared to the He~I line,
can appear earlier in type due to noise and, vice versa, a spectrum may be
classified as a later type when the Mg~II line is stronger than the He~I
line. In Fig.~\ref{spt:b} the distribution of spectral types in the 
sample is shown: most are in fact proposed to be mid B stars.

We show in Fig.~\ref{sn_fitting} some examples of our spectra within the 
4300--4500~\AA\ window compared with MS model atmospheres appropriate
to the chosen MS spectral sub-type. 

Luminosity class, for late-B and A stars, is in principle
well-determined from the appearance of the Balmer lines
(particularly the wings).  For B sub-types earlier than B4, \citet{Gray09}
cite relative strengths of O~II and Si~II-IV absorption lines compared 
with H~I and He~I ones as luminosity-sensitive also.   Assigning the 
right luminosity class is much more difficult than assigning spectral
sub-type since emission in the Balmer series interferes with our view
of the Balmer line profiles for many of our objects.   Furthermore the 
combination of S/N ratio and moderate spectral resolution reduces the 
possibility to classify using the Balmer-line wings and renders the 
weaker O and Si gravity-sensitive transitions undetectable. An
evaluation of the class III-V uncertainty and its impact on the
distance determination will be discussed in Section~\ref{chap4.2}
and \ref{chap5.1.1}.
 
The spectral types assigned to the observed stars are set out in 
Table~\ref{reddenings_t} where, for the moment, the luminosity class
is left unassigned.

\setcounter{figure}{5}
\begin{figure*}
\centering
\includegraphics[width=0.8\linewidth]{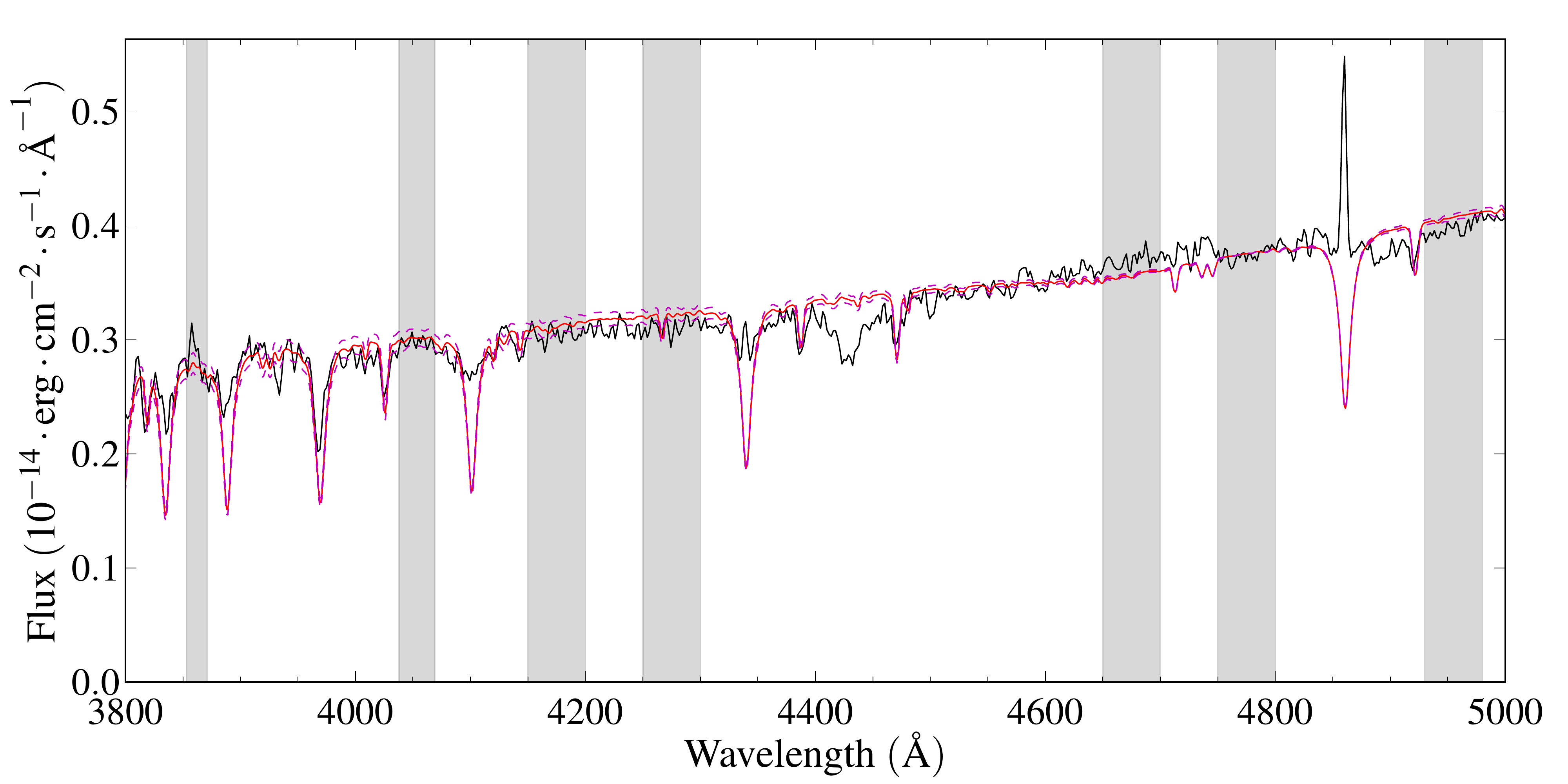}
\caption{Example of a reddening measurement based on a blue
  spectrum. The INT spectrum of object \#16 (black line) is shown along
with the model atmosphere for $T_{\rm{eff}} = 17000K$, reddened according
to the best fitting colour excess ($E(B-V)_{{\rm S}} = 1.27\pm0.04$).  The 
reddened model is drawn in red, with its 1-$\sigma$ error bounds shown as 
purple dashed lines.  The shaded vertical strips pick out the continuum 
intervals used in the fitting procedure. The normalisation applied in this 
instance is at $\lambda \, 4775 \rm{\AA}$. Note that the poor agreement between the
template Balmer line profiles and those of the star is due to infilling line
emission.}
\label{fits}
\end{figure*}
\subsection{Reddenings}
\label{chap3.2}

Two methods are used to measure the reddening of each star in the
sample.  The first, our primary method that we deploy in the later
parts of this study, is spectrophotometric and should be very
sensitive since we access the blue part of the spectrum 
(3800--5000~\AA), for all the objects.  The second is essentially 
photometric, in that it makes use of the IPHAS $(r-i)$ colour but
requires knowledge of spectral type (supplied by the spectroscopy).
Given the presence of circumstellar excess emission, which is
wavelength dependent, we expect to see a difference between the two 
determinations, in the sense that the photometric value is greater.
We compute this second reddening to see if this expectation is borne
out.

\subsubsection{Reddening estimation: spectroscopic method}
\label{chap3.2.1}

A least-squares fitting method was applied as follows.

First, we map the spectral sub-types of Section~\ref{chap3.1} onto an
approximate $T_{\rm{eff}}$ scale, using \citet{Kenyon95}
for main sequence stars (see Table~\ref{intrinsic}). 
Then, the basic idea of the fit is to compare each observed spectrum 
with the corresponding solar-abundance model for the
appropriate $T_{\rm{eff}}$, with $\log (g) = 4.0$,
taken from the \citet{Munari05} library, 
as it is increasingly reddened -- thereby seeking out the minimum 
reduced $\chi^2$.  Numerical experiments show that the treatment of
all objects as class V stars, when they may be more luminous
class IV or III stars, introduces negligible error compared to all
other terms in the error budget (see below). 

So that the fitting is sensitive only to the overall slope of the
observed SED compared with its theoretical value and not to the
details of individual lines, the fits are carried out within carefully 
chosen spectral intervals that are free of structure due to 
deep absorption lines/bands (mainly the Balmer lines and DIBs).  In
effect we degrade both observation and model atmosphere to a number of
'line-free' narrow bands falling in the range 
$\lambda \lambda 3800$~--~$5000$~\AA.  Flux is averaged in each of these
fit intervals and weighted according to the measured noise.  In
the fitting software, the reference model is progressively
reddened, raising $E(B-V)_{{\rm S}}$ by 0.01~mag at each step, and the quality
of fit to the observed spectrum is appraised by calculating
$\chi^2$.  In this approach, the number of degrees of freedom,
$\nu$, is the number of adopted spectral intervals less the number of
free parameters -- here the latter number is 1 (for the reddening).  
In practice, fits were performed for two 
different normalisations of the model atmosphere to the data 
at 4250~\AA\, and 4750~\AA, with the final 
reddening being the average of the two slightly different outcomes. 

The reddening law used in all cases is based on the formulation
given in \citet{Fitzpatrick99} with  $R_V = 3.1$. The choice of $R_V$ 
to within a few tenths has little impact on the derived colour excess,
as small changes in $R_V$ scarcely change the slope of the lav in the 
blue-visual range. Nevertheless it does 
affect the distance estimates as we will explain in Section~\ref{chap5}.
One example of the results of the fit
process is displayed, along with the selected wavelength intervals
used in the fits, in Fig.~\ref{fits}.

Errors on $E(B-V)_{{\rm S}}$ are determined graphically, by
identifying the
$\Delta\chi^2 \leq 1$ range around the minimum. We find that these are
typically $\pm 0.05$ magnitudes.

In principle a systematic error is introduced into the determination
of $E(B-V)_{{\rm S}}$, if the spectral type and mapping onto a reference model 
atmosphere are incorrect.   Since the Planck maximum in B and even
early-A stars is in the ultraviolet, their SEDs are tending towards
the Rayleigh-Jeans limit in the optical.  As a consequence the
spectral type uncertainty does not generate a large extra error
in $E(B-V)_{{\rm S}}$.  Experiments in which the adopted model
atmosphere is altered by $\pm 1$ sub-type or uprated to luminosity
class III, indicate a further error of up to 
$\pm 0.05$~mag in $E(B-V)_{{\rm S}}$. There is, in
addition, a random component linked to the known SED/colour spread 
associated with any one spectral type: based on    
the {\it Hipparcos} dataset \citet{Houk97} showed, 
for B8~--~F3 stars, $\sigma (B-V) \sim 0.03$. %The choice of 
In the error budget, therefore, the direct fit error is in average equal or 
larger than the other sources of uncertainty.

The measured spectroscopic reddenings, $E(B-V)_{\rm{S}}$, are listed in 
Table~\ref{reddenings_t}. 
\setcounter{table}{3}
\begin{table}
\caption[]{Adopted class V $T_{\rm{eff}}$ scale, intrinsic colours and absolute magnitude scale. The $T_{\rm{eff}}$ values are from \citet{Kenyon95}; the intrinsic colours are computed taking the average of 
\citet{Sale09}, Fabregat (priv. comm.), \citet{Kenyon95}, \citet{Siess97}. The absolute $r$ magnitudes are conversions of the absolute $V$ magnitudes given by
\citet{ZB91}. In the final two columns we give class IV and III absolute magnitudes obtained from the same source.  Uncertainties on $Mr$ are $50 \%$ of the absolute 
errors given by \citet{ZB91}, which more closely resemble the standard deviations at each sub-type than the full range specified by \citet{ZB91}.}
\centering
\begin{tabular}{@{}lrrrrr@{}}
\hline
    &        \multicolumn{3}{c}{dwarfs}  &     \multicolumn{1}{r}{subgiants}  &     \multicolumn{1}{r}{giants}      \\
SpT & $T_{\rm{eff}}$  &$(r - i)_{\rm{o}}$ &  $\rm{M}_{r}$&  $\rm{M}_{r}$&  $\rm{M}_{r}$ \\
 &    (K) &(mag) &  (mag)&  (mag)&  (mag) \\
\hline
  B0 &30000&$-0.17 $&$-3.40 \pm 0.30 $&$-3.70 \pm 0.25 $&$-4.20 \pm 0.35$\\
  B1 &25400&$-0.15 $&$-2.80 \pm 0.30 $&$-3.10 \pm 0.20 $&$-3.70 \pm 0.35$\\
  B2 &22000&$-0.13 $&$-2.10 \pm 0.35 $&$-2.50 \pm 0.30 $&$-3.30 \pm 0.45$\\
  B3 &18700&$-0.12 $&$-1.55 \pm 0.25 $&$-2.00 \pm 0.20 $&$-2.85 \pm 0.50$\\
  B4 &17000&$-0.09 $&$-1.15 \pm 0.20 $&$-1.65 \pm 0.20 $&$-2.45 \pm 0.55$\\
  B5 &15400&$-0.08 $&$-0.70 \pm 0.20 $&$-1.20 \pm 0.25 $&$-2.30 \pm 0.55$\\
  B6 &14000&$-0.07 $&$-0.30 \pm 0.20 $&$-0.75 \pm 0.25 $&$-1.90 \pm 0.55$\\
  B7 &13000&$-0.06 $&$-0.10 \pm 0.20 $&$-0.50 \pm 0.25 $&$-1.60 \pm 0.50$\\
  B8 &11900&$-0.04 $&$ 0.20 \pm 0.25 $&$-0.30 \pm 0.25 $&$-1.30 \pm 0.50$\\
  B9 &10500&$-0.02 $&$ 0.60 \pm 0.25 $&$ 0.10 \pm 0.30 $&$-0.90 \pm 0.60$\\
  A0 & 9520&$ 0.00 $&$ 1.00 \pm 0.25 $&$ 0.50 \pm 0.30 $&$-0.50 \pm 0.60$\label{intrinsic}\\
\hline
\end{tabular}
\end{table}
\setcounter{table}{2}
\begin{table*}
\begin{minipage}{130mm}
\caption[]{Spectral parameters of the 67 CBe stars, as derived in Section~\ref{chap3}. 
Columns are in the following order: ID number; spectral type; S/N at $\lambda\,
4500$~\AA; measured colour excess; 
H$\alpha$ emission-corrected $(r - i)_{\rm{c}}$ colours; photometric colour excess; 
absorption-corrected $\Halpha$ equivalent width; disc fraction from the scaling relation, 
equation~\eqref{dachs:b}. The final columns lists the spectroscopic interstellar reddening 
$E(B-V)_{\rm{(S, c)}}$, after correction for the circumstellar excess, and the asymptotic 
value of $E(B-V)$ for the sight-line from \citetalias{SFD98}.}
\begin{tabular}{@{}r l r r r r r r r c @{}}
\hline
\# & SpT &S/N & $E(B-V)_{\rm{S}}$ & $(r-i)_{\rm{c}}$ &  $E(B-V)_{\rm{P}}$& $EW(\Halpha)$ &  $f_{\rm{D}}$&  $E(B-V)_{\rm{(S, c)}}$&  $E(B-V)_{\rm{SFD98}}$  \\
   &     & &         (mag)               &    (mag)              &     (mag)             &(\rm{\AA}) \label{reddenings_t}        &  &(mag)&(mag)\\
\hline
$1 $& B5    &	47&$ 1.40 \pm 0.08 $&$ 0.87 $&$ 1.37 \pm 0.09 $&$ -25.4 \pm 1.1 $&$ 0.08  $&$ 1.36 \pm 0.08 $&$ 1.51$\\
$2 $& B7    &	67&$ 0.66 \pm 0.07 $&$ 0.35 $&$ 0.60 \pm 0.05 $&$ -12.6 \pm 0.9 $&$ 0.04  $&$ 0.64 \pm 0.07 $&$ 1.58$\\
$3 $& B3    &	40&$ 1.60 \pm 0.08 $&$ 1.00 $&$ 1.62 \pm 0.10 $&$ -34.6 \pm 0.8 $&$ 0.12  $&$ 1.54 \pm 0.08 $&$ 1.83$\\
$4 $& A0    &	29&$ 1.02 \pm 0.09 $&$ 0.77 $&$ 1.12 \pm 0.08 $&$ -22.5 \pm 1.3 $&$ 0.08  $&$ 0.98 \pm 0.09 $&$ 1.78$\\
$5 $& B2    &	44&$ 1.14 \pm 0.08 $&$ 0.73 $&$ 1.25 \pm 0.08 $&$ -44.7 \pm 0.9 $&$ 0.15  $&$ 1.07 \pm 0.08 $&$ 1.55$\\
$6 $& B3    &	25&$ 1.38 \pm 0.10 $&$ 0.91 $&$ 1.49 \pm 0.09 $&$ -25.8 \pm 1.1 $&$ 0.09  $&$ 1.34 \pm 0.10 $&$ 1.51$\\
$7 $& B7    &	35&$ 0.84 \pm 0.07 $&$ 0.51 $&$ 0.82 \pm 0.06 $&$ -17.5 \pm 1.2 $&$ 0.06  $&$ 0.81 \pm 0.07 $&$ 1.51$\\
$8 $& B3    &	31&$ 1.12 \pm 0.07 $&$ 0.63 $&$ 1.09 \pm 0.07 $&$ -30.2 \pm 1.4 $&$ 0.10  $&$ 1.07 \pm 0.07 $&$ 1.28$\\
$9 $& B5    &	37&$ 0.94 \pm 0.08 $&$ 0.56 $&$ 0.93 \pm 0.06 $&$ -17.6 \pm 1.0 $&$ 0.06  $&$ 0.91 \pm 0.08 $&$ 1.60$\\
$10 $& B7   &	40&$ 1.10 \pm 0.08 $&$ 0.68 $&$ 1.08 \pm 0.07 $&$ -33.8 \pm 0.8 $&$ 0.11  $&$ 1.05 \pm 0.08 $&$ 1.66$\\
$11 $& B2-3 &	55&$ 0.96 \pm 0.08 $&$ 0.65 $&$ 1.12 \pm 0.07 $&$ -48.5 \pm 0.9 $&$ 0.16  $&$ 0.88 \pm 0.08 $&$ 1.07$\\
$12 $& B5   &	49&$ 0.86 \pm 0.09 $&$ 0.51 $&$ 0.85 \pm 0.06 $&$ -19.6 \pm 1.2 $&$ 0.07  $&$ 0.83 \pm 0.09 $&$ 1.15$\\
$13 $& B5   &	81&$ 0.66 \pm 0.08 $&$ 0.38 $&$ 0.67 \pm 0.05 $&$ -13.6 \pm 1.0 $&$ 0.05  $&$ 0.64 \pm 0.08 $&$ 1.02$\\
$14 $& B4   &	61&$ 1.14 \pm 0.08 $&$ 0.75 $&$ 1.22 \pm 0.08 $&$ -31.4 \pm 1.3 $&$ 0.10  $&$ 1.09 \pm 0.08 $&$ 1.37$\\
$15 $& B5   &	87&$ 1.10 \pm 0.07 $&$ 0.74 $&$ 1.18 \pm 0.08 $&$ -25.4 \pm 0.8 $&$ 0.08  $&$ 1.06 \pm 0.07 $&$ 1.66$\\
$16 $& B3   &	67&$ 1.27 \pm 0.08 $&$ 0.88 $&$ 1.45 \pm 0.09 $&$ -29.1 \pm 0.8 $&$ 0.10  $&$ 1.22 \pm 0.08 $&$ 1.91$\\
$17 $& B3   &	38&$ 1.53 \pm 0.08 $&$ 1.01 $&$ 1.64 \pm 0.10 $&$ -91.6 \pm 0.9 $&$ 0.31  $&$ 1.36 \pm 0.08 $&$ 1.47$\\
$18 $& B4   &	54&$ 1.07 \pm 0.08 $&$ 0.74 $&$ 1.20 \pm 0.08 $&$ -48.1 \pm 1.0 $&$ 0.16  $&$ 0.99 \pm 0.08 $&$ 1.39$\\
$19 $& B6   &	28&$ 1.14 \pm 0.07 $&$ 0.64 $&$ 1.03 \pm 0.07 $&$ -23.8 \pm 1.2 $&$ 0.08  $&$ 1.10 \pm 0.07 $&$ 1.19$\\
$20 $& B3   &	51&$ 1.40 \pm 0.08 $&$ 0.99 $&$ 1.61 \pm 0.10 $&$ -72.6 \pm 0.8 $&$ 0.24  $&$ 1.28 \pm 0.08 $&$ 1.90$\\
$21 $& B5   &	30&$ 1.40 \pm 0.09 $&$ 0.87 $&$ 1.38 \pm 0.09 $&$ -43.8 \pm 1.3 $&$ 0.15  $&$ 1.33 \pm 0.09 $&$ 2.39$\\
$22 $& B4   &	41&$ 1.33 \pm 0.07 $&$ 0.92 $&$ 1.47 \pm 0.09 $&$ -27.3 \pm 1.0 $&$ 0.09  $&$ 1.29 \pm 0.07 $&$ 1.40$\\
$23 $& B7   &	48&$ 1.08 \pm 0.09 $&$ 0.77 $&$ 1.20 \pm 0.08 $&$ -50.0 \pm 1.0 $&$ 0.17  $&$ 1.00 \pm 0.09 $&$ 1.39$\\
$24 $& B3   &	54&$ 1.36 \pm 0.07 $&$ 0.89 $&$ 1.46 \pm 0.09 $&$ -32.2 \pm 0.8 $&$ 0.11  $&$ 1.31 \pm 0.07 $&$ 1.96$\\
$25 $& B5   &	79&$ 0.86 \pm 0.07 $&$ 0.55 $&$ 0.92 \pm 0.06 $&$ -25.0 \pm 0.8 $&$ 0.08  $&$ 0.82 \pm 0.07 $&$ 1.19$\\
$26 $& B3   &	33&$ 1.18 \pm 0.09 $&$ 0.85 $&$ 1.41 \pm 0.09 $&$ -99.4 \pm 1.1 $&$ 0.33  $&$ 1.00 \pm 0.09 $&$ 1.37$\\
$27 $& B5   &	44&$ 1.08 \pm 0.07 $&$ 0.67 $&$ 1.09 \pm 0.07 $&$ -19.6 \pm 1.0 $&$ 0.07  $&$ 1.05 \pm 0.07 $&$ 1.30$\\
$28 $& B7   &	50&$ 0.83 \pm 0.08 $&$ 0.48 $&$ 0.78 \pm 0.06 $&$ -20.0 \pm 0.9 $&$ 0.07  $&$ 0.80 \pm 0.08 $&$ 1.12$\\
$29 $& B7   &	69&$ 0.80 \pm 0.07 $&$ 0.50 $&$ 0.81 \pm 0.06 $&$ -19.4 \pm 0.9 $&$ 0.06  $&$ 0.77 \pm 0.07 $&$ 1.42$\\
$30 $& B4   &	45&$ 0.78 \pm 0.07 $&$ 0.45 $&$ 0.78 \pm 0.06 $&$ -5.3 \pm 1.2 $&$ 0.02  $&$ 0.77 \pm 0.07 $&$ 0.97$\\
$31 $& B3   &	56&$ 1.28 \pm 0.07 $&$ 0.89 $&$ 1.46 \pm 0.09 $&$ -24.5 \pm 0.9 $&$ 0.08  $&$ 1.24 \pm 0.07 $&$ 1.74$\\
$32 $& B6   &	66&$ 1.01 \pm 0.08 $&$ 0.54 $&$ 0.88 \pm 0.06 $&$ -25.6 \pm 0.9 $&$ 0.09  $&$ 0.97 \pm 0.08 $&$ 0.93$\\
$33 $& B8-9 &	77&$ 0.88 \pm 0.07 $&$ 0.58 $&$ 0.88 \pm 0.07 $&$ -14.4 \pm 0.7 $&$ 0.05  $&$ 0.86 \pm 0.07 $&$ 1.02$\\
$34 $& B3   &	47&$ 0.70 \pm 0.08 $&$ 0.49 $&$ 0.88 \pm 0.06 $&$ -36.9 \pm 0.9 $&$ 0.12  $&$ 0.64 \pm 0.08 $&$ 1.18$\\
$35 $& B2-3 &	47&$ 0.93 \pm 0.08 $&$ 0.58 $&$ 1.02 \pm 0.07 $&$ -56.5 \pm 1.1 $&$ 0.19  $&$ 0.84 \pm 0.08 $&$ 1.09$\\
$36 $& B4   &	37&$ 1.53 \pm 0.10 $&$ 1.11 $&$ 1.74 \pm 0.11 $&$ -59.0 \pm 1.0 $&$ 0.20  $&$ 1.43 \pm 0.10 $&$ 2.06$\\
$37 $& B6   &	51&$ 0.92 \pm 0.08 $&$ 0.61 $&$ 0.98 \pm 0.07 $&$ -18.5 \pm 0.9 $&$ 0.06  $&$ 0.89 \pm 0.08 $&$ 1.19$\\
$38 $& B5   &	22&$ 0.90 \pm 0.15 $&$ 0.68 $&$ 1.10 \pm 0.07 $&$ -21.0 \pm 1.3 $&$ 0.07  $&$ 0.87 \pm 0.15 $&$ 1.61$\\
$39 $& B4   &	27&$ 0.96 \pm 0.12 $&$ 0.75 $&$ 1.22 \pm 0.08 $&$ -82.4 \pm 1.1 $&$ 0.27  $&$ 0.82 \pm 0.12 $&$ 0.93$\\
$*40 $& B2  &	 52&$ 1.04 \pm 0.07 $&$ 0.67 $&$ 1.16 \pm 0.07 $&$ -13.2 \pm 1.5 $&$ 0.04  $&$ 1.02 \pm 0.07 $&$ 0.32$\\
$*41 $& B7  &	 64&$ 1.02 \pm 0.07 $&$ 0.57 $&$ 0.92 \pm 0.06 $&$ -19.2 \pm 1.6 $&$ 0.06  $&$ 0.99 \pm 0.07 $&$ 0.95$\\
$*42 $& B2  &	 22&$ 1.47 \pm 0.08 $&$ 1.00 $&$ 1.63 \pm 0.10 $&$ -70.3 \pm 2.2 $&$ 0.23  $&$ 1.36 \pm 0.08 $&$ 0.94$\\
$*43 $& B6  &	 48&$ 0.82 \pm 0.07 $&$ 0.44 $&$ 0.74 \pm 0.05 $&$		  $&$		  $&$ 0.82 \pm 0.07 $&$ 0.68$\\
$*44 $& B5  &	 51&$ 0.79 \pm 0.07 $&$ 0.56 $&$ 0.93 \pm 0.06 $&$ -9.6 \pm 2.6 $&$ 0.03  $&$ 0.78 \pm 0.07 $&$ 0.85$\\
$45 $& B3   &	103&$ 1.14 \pm 0.07 $&$ 0.64 $&$ 1.10 \pm 0.07 $&$ -19.2 \pm 0.8 $&$ 0.06  $&$ 1.11 \pm 0.07 $&$ 1.51$\\
$*46 $& B9  &	 40&$ 0.84 \pm 0.07 $&$ 0.57 $&$ 0.86 \pm 0.07 $&$ -16.8 \pm 1.6 $&$ 0.06  $&$ 0.81 \pm 0.07 $&$ 1.19$\\
$*47 $& B3  &	 36&$ 1.65 \pm 0.07 $&$ 1.13 $&$ 1.81 \pm 0.11 $&$ -42.2 \pm 1.2 $&$ 0.14  $&$ 1.58 \pm 0.07 $&$ 1.98$\\
$*48 $& B8  &	 22&$ 1.15 \pm 0.10 $&$ 0.74 $&$ 1.13 \pm 0.08 $&$		  $&$		  $&$ 1.15 \pm 0.10 $&$ 1.23$\\
$*49 $& A0  &	 43&$ 0.74 \pm 0.07 $&$ 0.44 $&$ 0.64 \pm 0.05 $&$ -26.5 \pm 3.3 $&$ 0.09  $&$ 0.70 \pm 0.07 $&$ 0.72$\\
$*50 $& B3  &	 34&$ 1.00 \pm 0.08 $&$ 0.75 $&$ 1.26 \pm 0.08 $&$ -47.0 \pm 2.0 $&$ 0.16  $&$ 0.92 \pm 0.08 $&$ 1.26$\\
$*51 $& B8-9&	 63&$ 0.73 \pm 0.07 $&$ 0.41 $&$ 0.63 \pm 0.06 $&$ -12.3 \pm 1.3 $&$ 0.04  $&$ 0.71 \pm 0.07 $&$ 0.78$\\
$*52 $& B7  &	 40&$ 0.90 \pm 0.08 $&$ 0.60 $&$ 0.96 \pm 0.07 $&$ -18.0 \pm 2.7 $&$ 0.06  $&$ 0.87 \pm 0.08 $&$ 0.98$\\
$*53 $& B7  &	 45&$ 1.11 \pm 0.07 $&$ 0.67 $&$ 1.05 \pm 0.07 $&$ -41.6 \pm 1.7 $&$ 0.14  $&$ 1.04 \pm 0.07 $&$ 1.47$\\
$*54 $& B3  &	 51&$ 0.85 \pm 0.07 $&$ 0.48 $&$ 0.86 \pm 0.06 $&$ -26.8 \pm 2.0 $&$ 0.09  $&$ 0.81 \pm 0.07 $&$ 1.12$\\
$*55 $& B3-4&	 28&$ 1.04 \pm 0.08 $&$ 0.72 $&$ 1.19 \pm 0.08 $&$ -37.8 \pm 2.0 $&$ 0.13  $&$ 0.98 \pm 0.08 $&$ 1.22$\\
\hline
\multicolumn{10}{p{\linewidth}}{Note: $^{*}$ NOT/ALFOSC observations, for which $\Halpha$ equivalent widths were measured from FLWO-1.5m/FAST spectra when available.}
\end{tabular}
\end{minipage}
\end{table*}

\begin{table*}
\begin{minipage}{130mm}
\contcaption{}
\begin{tabular}{@{}r l r r r r r r r c@{}}
\hline
\# & SpT & S/N &$E(B-V)_{\rm{S}}$ &  $(r-i)_{\rm{c}}$ &  $E(B-V)_{\rm{P}}$& $EW(\Halpha)$ &  $f_{\rm{D}}$&  $E(B-V)_{\rm{(S, c)}}$&  $E(B-V)_{\rm{SFD98}}$  \\
   &     &   &        (mag)               &    (mag)              &     (mag)             &(\rm{\AA})         &  &(mag)&(mag)\\
\hline
$*56 $& B7   & 45&$ 1.02 \pm 0.07 $&$ 0.69 $&$ 1.09 \pm 0.07 $&$ -47.4 \pm 1.6 $&$ 0.16  $&$ 0.94 \pm 0.07 $&$ 1.72$\\
$*57 $& B7-8 & 52&$ 1.00 \pm 0.08 $&$ 0.70 $&$ 1.09 \pm 0.08 $&$	     $&$	     $&$ 1.00 \pm 0.08 $&$ 1.36$\\
$*58 $& B3   & 44&$ 0.72 \pm 0.07 $&$ 0.36 $&$ 0.70 \pm 0.05 $&$ -31.4 \pm 1.2 $&$ 0.10  $&$ 0.67 \pm 0.07 $&$ 0.90$\\
%$*59 $& B4   & 29&$ 1.04 \pm 0.09 $&$ 0.69 $&$ 1.13 \pm 0.08 $&$ -144.0 \pm 1.9 $&$ 0.48  $&$ 0.74 \pm 0.09 $&$ 3.16$\\
$*59 $& B5   & 44&$ 0.78 \pm 0.08 $&$ 0.47 $&$ 0.80 \pm 0.05 $&$ -52.2 \pm 1.8 $&$ 0.17  $&$ 0.70 \pm 0.08 $&$ 1.24$\\
$*60 $& B3-4 & 36&$ 0.82 \pm 0.08 $&$ 0.39 $&$ 0.72 \pm 0.06 $&$ -31.2 \pm 3.1 $&$ 0.10  $&$ 0.77 \pm 0.08 $&$ 1.24$\\
$*61 $& B7   & 51&$ 0.74 \pm 0.08 $&$ 0.43 $&$ 0.71 \pm 0.05 $&$ -11.4 \pm 1.6 $&$ 0.04  $&$ 0.72 \pm 0.08 $&$ 0.81$\\
$*62 $& B6   & 26&$ 1.23 \pm 0.10 $&$ 0.77 $&$ 1.21 \pm 0.08 $&$ -16.7 \pm 1.6 $&$ 0.06  $&$ 1.20 \pm 0.10 $&$ 1.62$\\
$*63 $& B7   & 58&$ 0.89 \pm 0.07 $&$ 0.52 $&$ 0.84 \pm 0.06 $&$	     $&$	     $&$ 0.89 \pm 0.07 $&$ 1.67$\\
$*64 $& B5   & 37&$ 1.25 \pm 0.09 $&$ 0.81 $&$ 1.29 \pm 0.08 $&$ -17.0 \pm 2.1 $&$ 0.06  $&$ 1.22 \pm 0.09 $&$ 1.32$\\
$*65 $& B5   & 48&$ 0.98 \pm 0.07 $&$ 0.56 $&$ 0.92 \pm 0.06 $&$ -31.4 \pm 1.4 $&$ 0.10  $&$ 0.93 \pm 0.07 $&$ 1.55$\\
$*66 $& B4   & 31&$ 1.10 \pm 0.08 $&$ 0.64 $&$ 1.05 \pm 0.07 $&$ -25.5 \pm 1.4 $&$ 0.08  $&$ 1.06 \pm 0.08 $&$ 1.72$\\
$67 $& B4    & 35&$ 1.14 \pm 0.11 $&$ 0.81 $&$ 1.31 \pm 0.09 $&$ -18.2 \pm 0.9 $&$ 0.06  $&$ 1.11 \pm 0.11 $&$ 1.51$\\
\hline
\multicolumn{10}{p{\linewidth}}{Note: $^{*}$ NOT/ALFOSC observations, for which $\Halpha$ equivalent widths were measured from FLWO-1.5m/FAST spectra when available.}
\end{tabular}
\end{minipage}
\end{table*}
\subsubsection{Reddening estimation: photometric method}
\label{chap3.2.2}
IPHAS photometry provides an observed $(r - i)$ colour that can be used
in conjunction with the now known spectral type to give another
reddening estimate.  The procedure we adopted to do this has
three steps:
\begin{enumerate} 
\item The observed $(r-i)$ colour is corrected to zero $\Halpha$
  emission, by reference to the synthetic tracks given in
  \citet[Table~4,][]{Drew05}. This is a small
  correction, in the range 0.01~--~0.05 magnitudes. Corrected colours, 
  $(r-i)_{\rm{c}}$, are in table~\ref{reddenings_t}.

\item The colour excess for each object is then: 
\begin{equation}
E(r-i) = (r-i)_{\rm{c}} -(r-i)_{\rm{o}},
\end{equation}  
where $(r-i)_{\rm{o}}$ is the intrinsic colour, consistent with the spectral
type assigned in Section~\ref{chap3.1}.  The adopted intrinsic colours
are set out in Table~\ref{intrinsic}. 
\item The $(B-V)$ colour excess is then computed as: 
\begin{equation}
E(B - V)_{\rm{P}} = E(r-i) / 0.69,
\end{equation} 
adopting the same $R_v = 3.1$ reddening curve as applied in 
Section~\ref{chap3.2.1}.   
\end{enumerate}

Random photometric uncertainties in $r$ and $i$ for these relatively
bright objects are small -- not exceeding 0.01. Further uncertainties
to include are:
\begin{enumerate}
\item the spread in intrinsic colour, as commented on above in 
Section~\ref{chap3.2.1}.

\item the uncertainty originating from the $\pm 1$ sub-type error in the
spectral-typing. Across the B class this averages to $\pm 0.02$ mag.
As for the SED fitting, an uncertainty on the luminosity classes would introduce
a small $\pm 0.01$ mag error. 

\end{enumerate}

Photometric reddenings, $E(B-V)_{\rm{P}}$, are also recorded in Table~\ref{reddenings_t}.
\setcounter{table}{4}
\begin{table*}
\begin{minipage}{130mm}
\caption[]{Circumstellar colour excess and $r$ magnitude corrections for a 
given spectral type and disc contribution to the total flux.}
\centering
\begin{tabular}{@{}l r c c c c c c c c @{}}
\hline
     &              &   \multicolumn{2}{c}{$f_{\rm{D}}$ = 0.05}  &  \multicolumn{2}{c}{$f_{\rm{D}}$ = 0.10} &
       \multicolumn{2}{c}{$f_{\rm{D}}$ = 0.20} &  \multicolumn{2}{c}{$f_{\rm{D}}$ = 0.30} \\
 SpT & $T_{e} (K)$  &        $E^{cs}(B-V)_{\rm{S}}$ & $\Delta r$ &$E^{cs}(B-V)_{\rm{S}}$ & $\Delta r$ &$E^{cs}(B-V)_{\rm{S}}$ & $\Delta r$ &$E^{cs}(B-V)_{\rm{S}}$ & $\Delta r$\\     
\hline
B1   & 18000 &   0.023 &  0.082 & 0.046  & 0.166   & 0.098  &  0.344  &  0.156  & 0.516  \\ 
B3   & 13200 &   0.022 &  0.085 & 0.048  & 0.173   & 0.103  &  0.357  &  0.164  & 0.534 \\  
B5   &  9300 &   0.024 &  0.089 & 0.049  & 0.180   & 0.105  &  0.369  &  0.169  & 0.552  \\  
B7   &  7800 &   0.024 &  0.093 & 0.049  & 0.188   & 0.105  &  0.385  &  0.170  & 0.573  \\  
A0   &  5700 &   0.023 &  0.104 & 0.047  & 0.209   & 0.103  &  0.424  &  0.168  & 0.627  \label{corrections}\\  
\hline
\end{tabular}
\end{minipage}
\end{table*}

\subsection{Correction for CBe circumstellar continuum emission}
\label{chap3.3}
\setcounter{figure}{6}
\begin{figure}
\includegraphics[width=\linewidth]{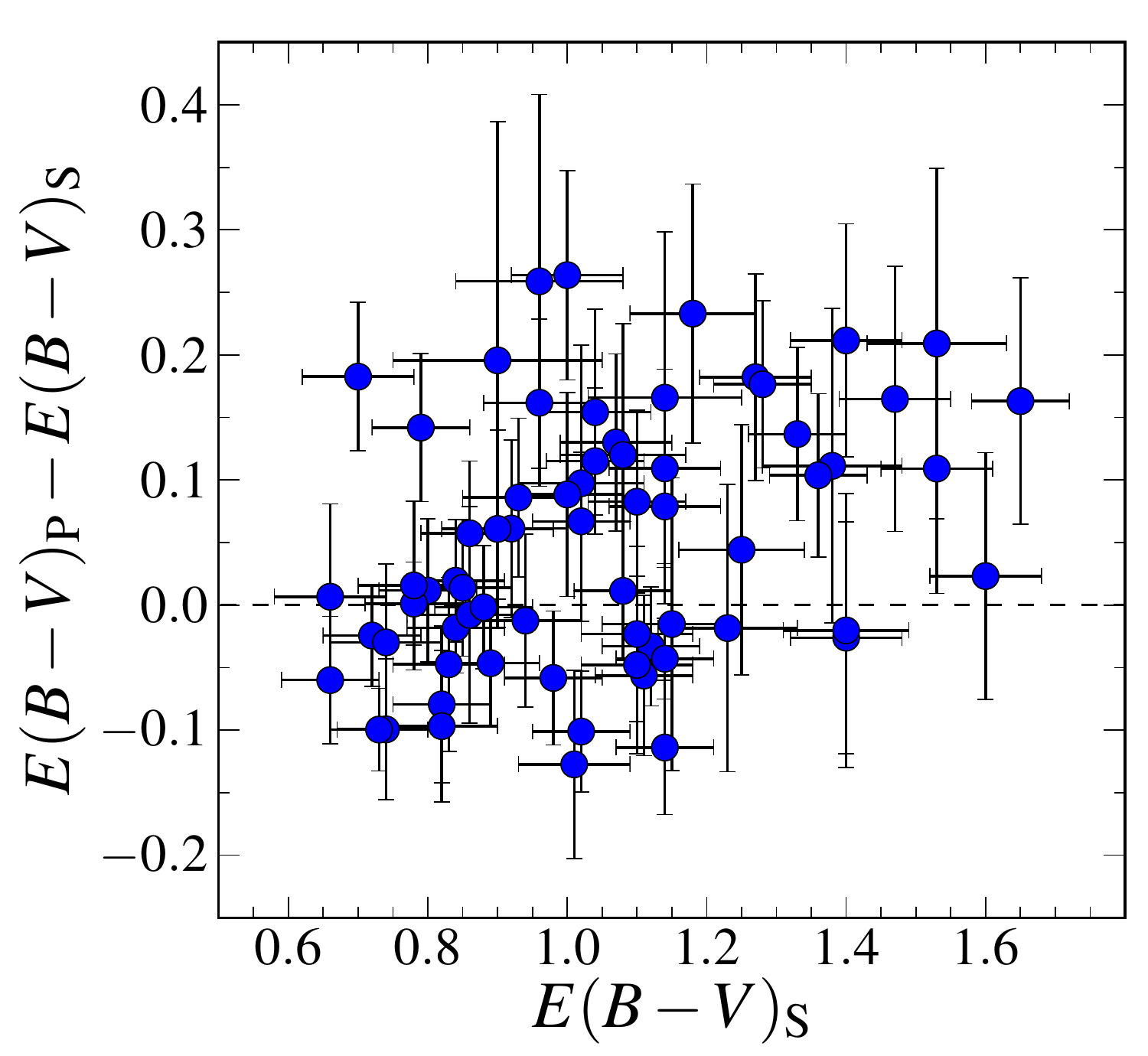}
\caption{The difference between the two colour excess measurements, 
$E(B-V)_{\rm{P}} - E(B-V)_{\rm{S}}$ is
  plotted as a function of the spectroscopic colour excess, $E(B-V)_{\rm {S}}$. 
The data points are scattered with a bias to positive values, as expected, due to the reddening 
effect associated with the circumstellar-disc emission present in these stars.}
\label{reddenings_f}
\end{figure}
\setcounter{figure}{7}
\begin{figure}
\includegraphics[width=\linewidth]{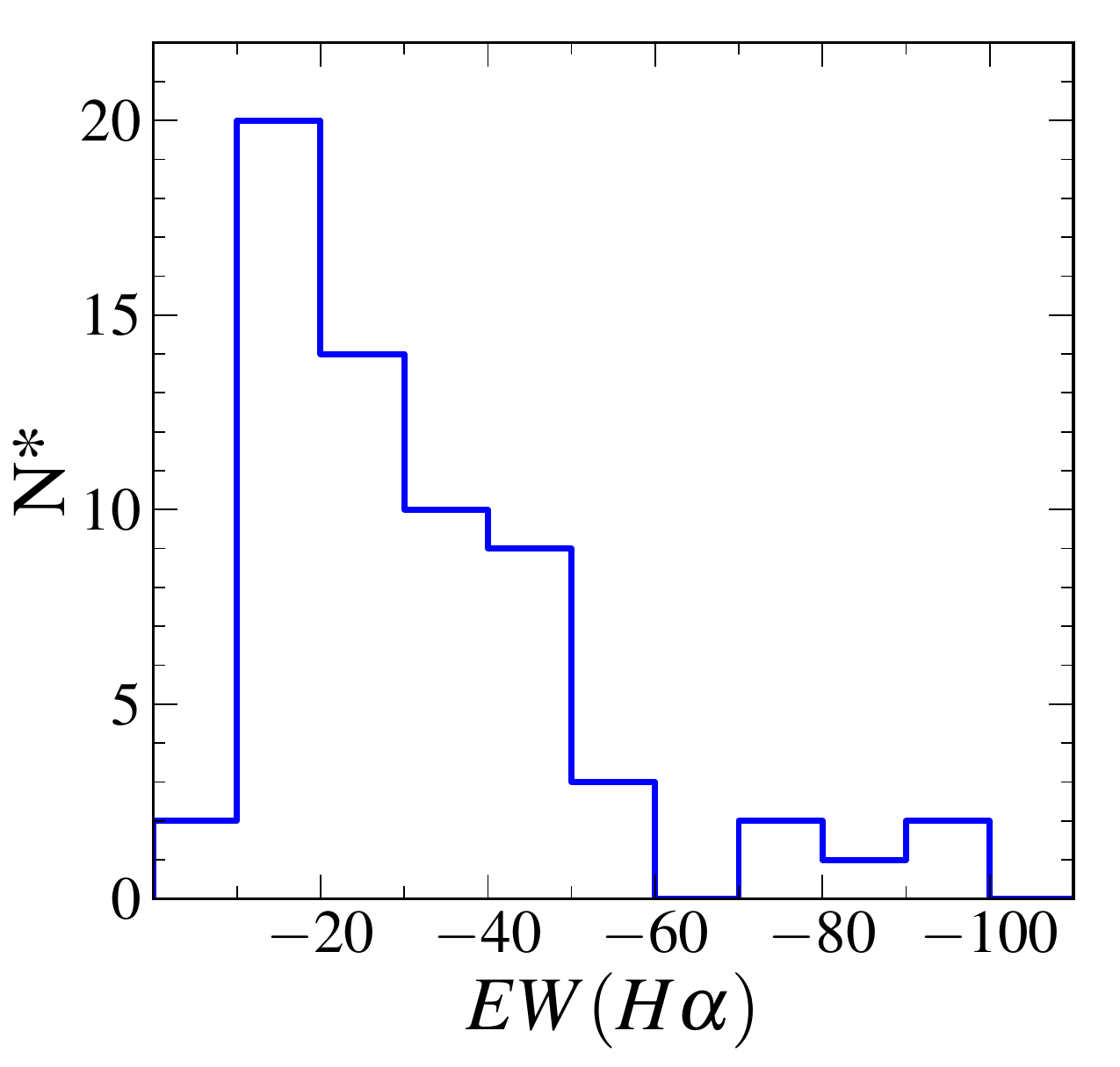}
\caption{The distribution of H$\alpha$ equivalent widths ($EW(\Halpha)$). 
The median of the distribution falls at $\approx -26$~\AA\, and all but 5 
of the 67 stars have $EW(\Halpha) \leq -60$~\AA.} 
\label{halpha_ews}
\end{figure}

CBe stars are affected by excess emission
which slightly alters the optical SED and induces an overestimate of the
colour excess, $E(B-V)$, if not taken into account. Following earlier notation \citep{Dachs88}, 
this component can be treated as additive to the interstellar value as in:
\begin{equation}
E(B-V) = E^{is}(B-V) + E^{cs}(B-V),
\end{equation}
where $E^{is}(B-V)$ is the interstellar reddening and $E^{cs}(B-V)$ is
the circumstellar contribution to the total colour excess.

\citet{Kaiser89} and, more recently \citet{Carciofi06}, have
demonstrated that the continuum excess, accounted for by $E^{cs}(B-V)$, 
can be attributed to an optically-thin free-free and recombination free-bound 
continuum . It is evident from this work that the wavelength
dependence of the disc continuum is such that the red spectrum
includes more disc light than the blue. \citet{Dachs88} 
specifically investigated the correlation between $EW(\Halpha)$ and 
$E^{cs}(B-V)$ and presented evidence that the former correlates
with the latter and also with the fraction of the total emission that 
can be attributed to the circumstellar disc. By analysing a sample of 
B0--B3 stars mainly, they found the following dependencies on $\Halpha$ 
emission equivalent width:

\begin{align}
E^{cs}(B-V) &\approx 0.02 \cdot \frac{EW(\Halpha)}{-10 \rm{\AA}}
\label{dachs:a}\\
f_{\rm D} = \frac{F_{\rm D}}{F_{{\rm D}}+ F^{*}} &\approx 0.1 \cdot \frac{EW(\Halpha)}{-30 \rm{\AA}}
\label{dachs:b},
\end{align}

where $f_{\rm{D}} = F_{\rm{D}}/(F_{\rm{D}}+ F^{*})$ is the fraction of flux emitted by the disc compared to the
total flux, at $\lambda\, 5500 \rm{\AA}$.

First we confirm that our sample of objects presents the expected
evidence of a continuum excess that affects the red-optical more than
the blue-optical.  Fig.~\ref{reddenings_f} compares the two
reddening measurements we have obtained for all members of the sample. 
In it, we notice a systematic overestimate of $E(B-V)_{\rm{P}}$ with
respect to $E(B-V)_{\rm{S}}$, which ties in with the description given by
\citet{Kaiser89}. Where $E(B-V)_{\rm P}$ is less than $E(B-V)_{\rm S}$ it
is never so negative that it may not be viewed as consistent with the
two measures being equal to within the errors.  This is encouraging
in the sense that this outcome would not be guaranteed if the sample
contained CBe stars prone to marked variability.  

A new feature of our sample compared to that of \citet{Dachs88} is 
that it includes 5 objects with $EW(\Halpha) \le -60 \rm{\AA}$
(see Fig.~\ref{halpha_ews}),  that therefore lie beyond the range
over which the correlations contained in equations~\eqref{dachs:a} and 
\eqref{dachs:b} were established.  

Our method for estimating the circumstellar colour excess begins with
equation~\eqref{dachs:b}, delivering the disc fraction, $f_{\rm{D}}$.  We
do not simply apply equation~\eqref{dachs:a} for the reason that it
was constructed to provide correction to reddenings measured directly
across the $B$ to $V$ range (roughly 4000 --- 6000~\AA ).  The
spectrophotometric reddening estimates obtained here are based on the
blue range only, stopping at 5000~\AA , where the contaminating 
circumstellar disc continuum will be less than the mean for the $B$
to $V$ range.  

We have computed some simple models that enable an appropriate scaling 
down of this correction.  It is assumed that the disc is optically thin 
at least in the Paschen continuum, emitting free-free and free-bound 
continuum emission from a fully ionised hydrogen envelope \citep{Dachs88, 
Kaiser89, Carciofi06}.  Our parametrisation is similar to that of 
\citet{Kaiser89}, in that we 
maintain the same definition of $f_{\rm{D}}$.  Our simulations cover the
range of spectral types present in our sample (B1 to A0), 
and disc fractions are varied from zero to a maximum of 0.45.  A significant
difference with respect to earlier treatments is that we adopt a scaling
of the electron temperature in the circumstellar disc such that 
$T_{e} = 0.6 \cdot T_{\rm{eff}}$: this has been shown to be a good approximation 
by \citet{Carciofi06} \citep[see also][]{Drew89}. The electron density is set at
the suitably high, representative value, 
$N_{e} = 10^{12}\, \rm{cm}^{-3}$ \citep{Dachs88, Dachs90}.

On this basis we generate the circumstellar continuum emission and add it to 
the \citet{Munari05} model atmospheres, scaling it as required 
at $\lambda\, 5500 \rm{\AA}$.  The correction, $E^{cs}(B-V)_{\rm{S}}$ can then
be determined by 'dereddening' the resultant total spectrum to match the model
atmosphere alone.  This is carried out within the wavelength range 
$\lambda \lambda\, 3800\rm{\AA}$~--~$5000\rm{\AA}$, paralleling the procedure
applied to the observed spectra (Section~\ref{chap3.2.1}). 

In Table~\ref{corrections} we provide a representative grid of spectral types 
and corresponding $E^{cs}(B-V)_{\rm{S}}$, for a given disc contribution ($f_{\rm{D}}$) to 
the total emitted flux. Later, the $r$ magnitudes of our sample will also need 
to be corrected to remove the circumstellar disc contribution. Our simulations
provide this correction, $\Delta r$, as well.  These are also given in 
Table~\ref{corrections}.  We find that the $r$ magnitudes of our sample 
will be brighter, due to circumstellar emission, by amounts ranging from zero 
up to 0.5 in the most extreme cases.

Since CBe stars are known to be erratic variables 
\citep[i.e.][]{ZB91, Porter03, Jones11}, we take
care to determine $f_{\rm D}$ from either observations of the
H$\alpha$ line that are simultaneous with our blue spectroscopy
(available with all our INT data), or from a well-validated proxy in 
the case of the NOT spectra without coverage of the H$\alpha$ region.
The necessary proxy is provided by the FLWO-1.5m/FAST spectra in which
we find that the H$\beta$ profile is a good match to that apparent in
the NOT spectrum.  Fortuitously there are good matches for all but
4 objects.  We list the values of $f_{\rm D}$ obtained for each
of our sample of stars in Table~\ref{reddenings_t}, where we also
give the H$\alpha$ emission equivalent width on which it is based. 
This quantity is corrected for the underlying absorption, according to 
spectral type \citep[see tabulation in][]{Jaschek87}.  The error on 
$f_{\rm{D}}$ mainly reflects the scatter in the original 
empirical relation due to \citet{Dachs88}.  We estimate this to be 
$\pm 0.02$ dex, and propagate it through into the $E^{cs}(B-V)$ error.

The final $E^{is}(B-V)$ is thus obtained by subtracting
our tailored estimate of $E^{cs}(B-V)$ from $E(B-V)_{\rm S}$.  This 
result is shown in the final column of Table~\ref{reddenings_t}.
The distribution of final corrected reddenings is displayed in Fig.~\ref{reddenings_histogram}.
It ranges between 0.6~--~1.5~mags. and the median $E(B-V)_{\rm{S,\,c}}$ is 0.98.
\setcounter{figure}{8}
\begin{figure}
\includegraphics[width=\linewidth]{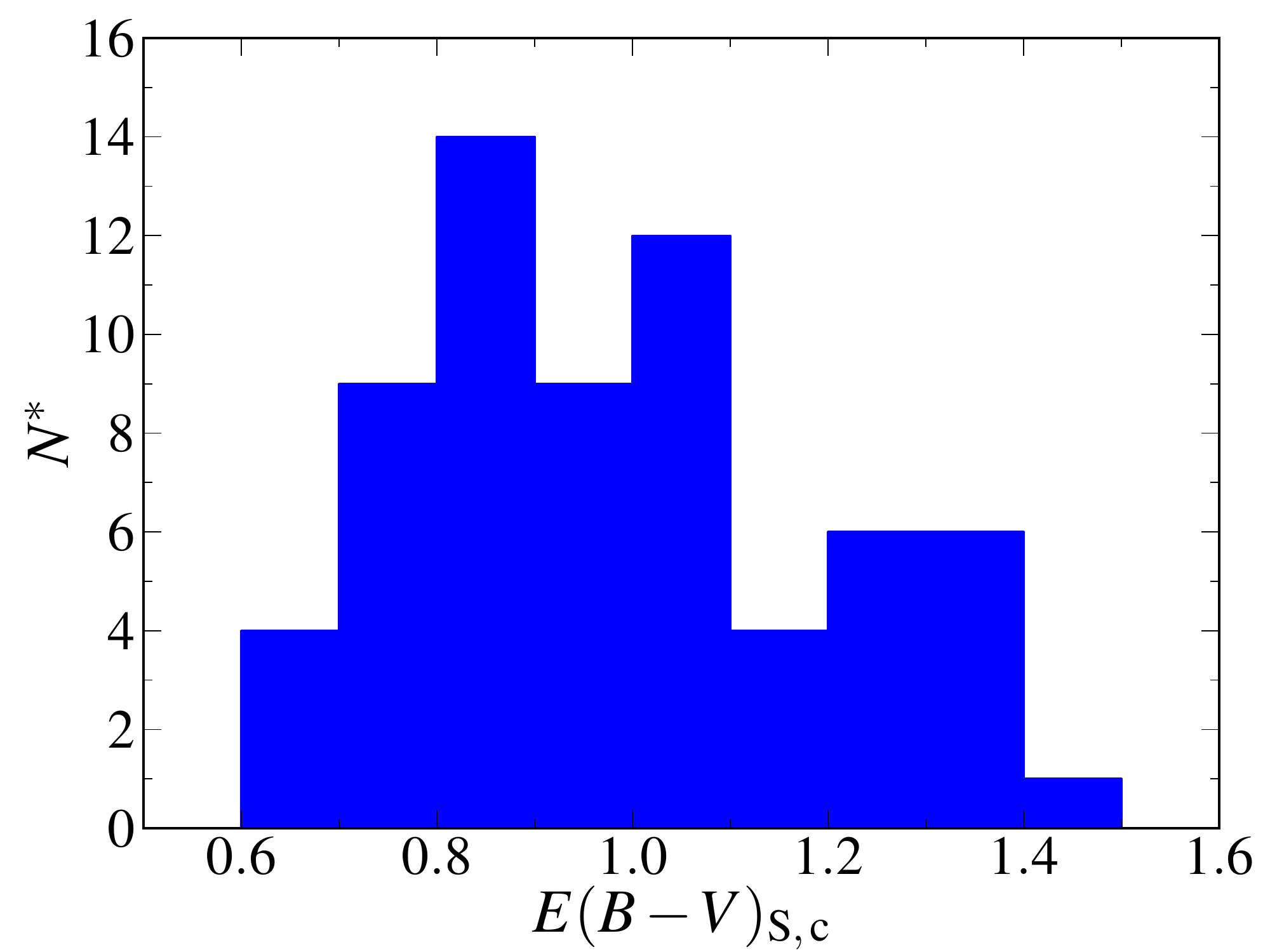}
\caption{The distribution of reddenings corrected for circumstellar colour excess. 
The median of the distribution is found at $E(B-V)_{\rm{S,\,c}} = 0.98$} 
\label{reddenings_histogram}
\end{figure}
\section{Distance estimation}
\label{chap4}
To set constraints on the distances to our objects, we first determine
spectroscopic parallaxes adopting absolute magnitudes of both luminosity 
class V, IV and III. Secondly, we constrain the luminosity class of the CBe stars
in the sample with help of IPHAS photometry of non-emission line stars that
are seen along the corresponding sightline and share similar reddenings with 
each CBe star.
\subsection{Distances from spectroscopic parallax}
\label{chap4.1}
Spectroscopic parallax distances, $D_{\rm{SP}}$, are computed in the standard 
way, via the use of spectral types and reddenings that were determined in 
Section~\ref{chap3} and the absolute magnitudes listed
in Table~\ref{intrinsic}. Our magnitude scale is taken from
\citet{ZB91} from which we also obtain error estimates;
if compared to others available in the literature \citep[e.g][]{Straizys81, Aller82, Wegner00} 
the Zorec \& Briot scale furnishes slightly fainter magnitudes than some although they agree
within the errors. We transformed their $V$-band absolute magnitudes
into $r$ absolute magnitudes, using the intrinsic $(V-R_{C})$ 
colours for dwarfs supplied by \citet{Kenyon95}, whilst noting that
$R_C$ and $r$ magnitudes of B stars in the Vega system are close enough to
identical for present purposes.  Furthermore, the differences between
dwarf and giant colours is small compared to all errors, permitting the
use of MS colours in obtaining $M_r$ for B giants. 

The observed $r$ magnitude needs to be corrected for the 
added flux due to circumstellar emission that makes the star look brighter than it would otherwise be (example
values for the correction, $\Delta r$, appear in Table~\ref{corrections}).
The extinction in the $r$ band is given by $A_{r} = 2.53 \cdot
E(B-V)_{\rm{(S, c)}}$, applying the same $R = 3.1$ extinction law
adopted in Section~\ref{chap3.2}. The main contributions to the
uncertainty in $D_{\rm{SP}}$ are the error in $E(B-V)_{\rm{(S, c)}}$
($\sigma \sim 0.1$) and in $M_{r}$ (as specified
in Table~\ref{intrinsic}).

In Table~\ref{t:distances} we list the
input corrected  magnitudes $r + \Delta r$ and $D_{\rm{SP}}$, computed
both for luminosity class V, IV and III. 

\subsection{Constraining the luminosity class}
\label{chap4.2}
In Section~\ref{chap3.1}, it was noted that the spectra used for typing are not of
the quality needed to pin down luminosity class. We now attempt to establish
some constraints on this by exploiting a very general property of the IPHAS 
colour-colour plane, which permits disentangling of intrinsic colours and 
reddenings of ordinary MS stars. To do this, we adapt the methods of 
analysis described in \cite{Drew08} and \cite{Sale10}, which focused on A-star
selection, and in \cite{Sale09}, which presented a more general 3D extinction 
mapping algorithm.  Essentially, for each CBe star, we pick out 
from IPHAS photometry fainter nearby non-emission line objects of similar 
reddening to see if, collectively, these putative lower main 
sequence companion objects favour a particular distance modulus, and hence 
-- by implication -- a particular luminosity class.  By this means we choose
between the class V, class IV and class III distance options listed in 
Table~\ref{t:distances}.

The method consists of the following steps and is illustrated by the 
two examples shown in Fig.~\ref{ms-fit}:
\setcounter{figure}{9}
\begin{figure*}
\centering
\subfigure[\# 11]{
\includegraphics[width=\linewidth]{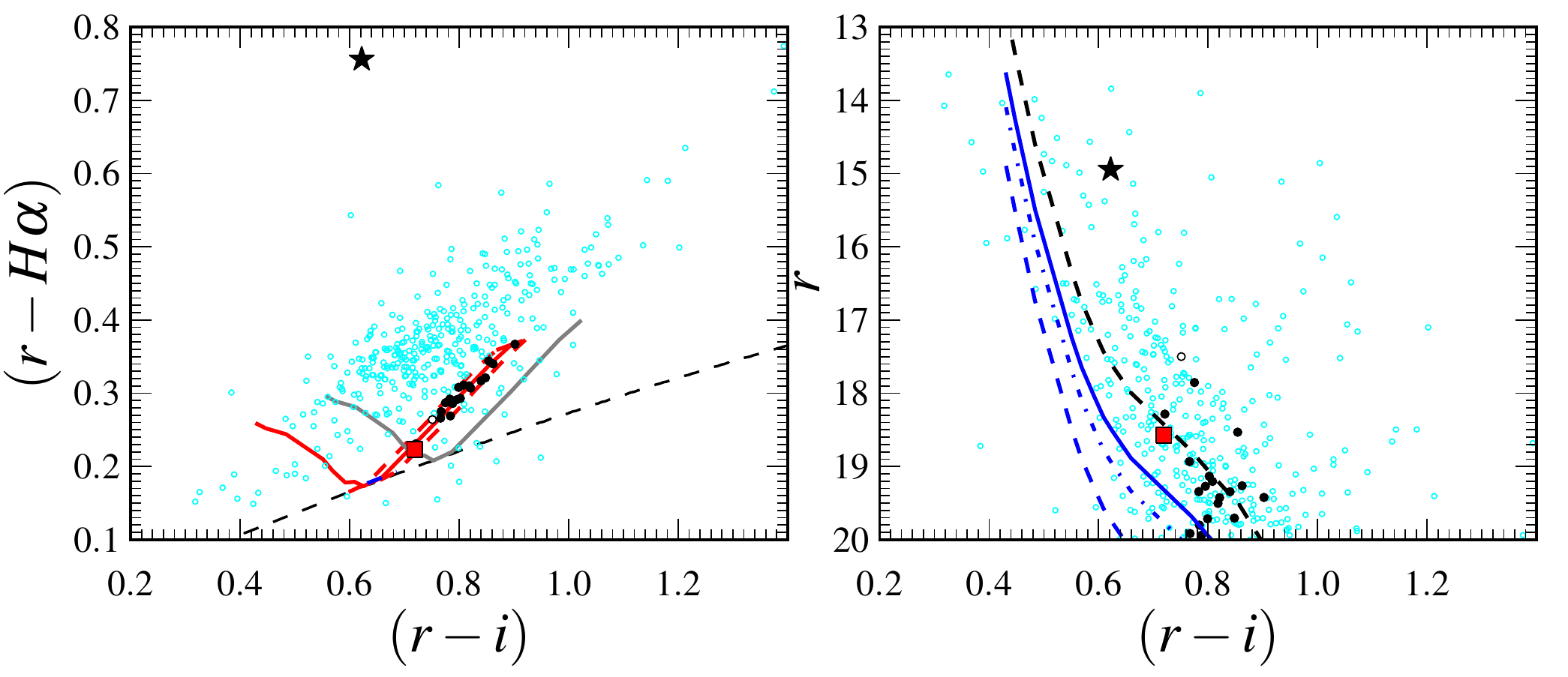}
\label{ms-fit:a}}
\subfigure[\# 52]{
\includegraphics[width=\linewidth]{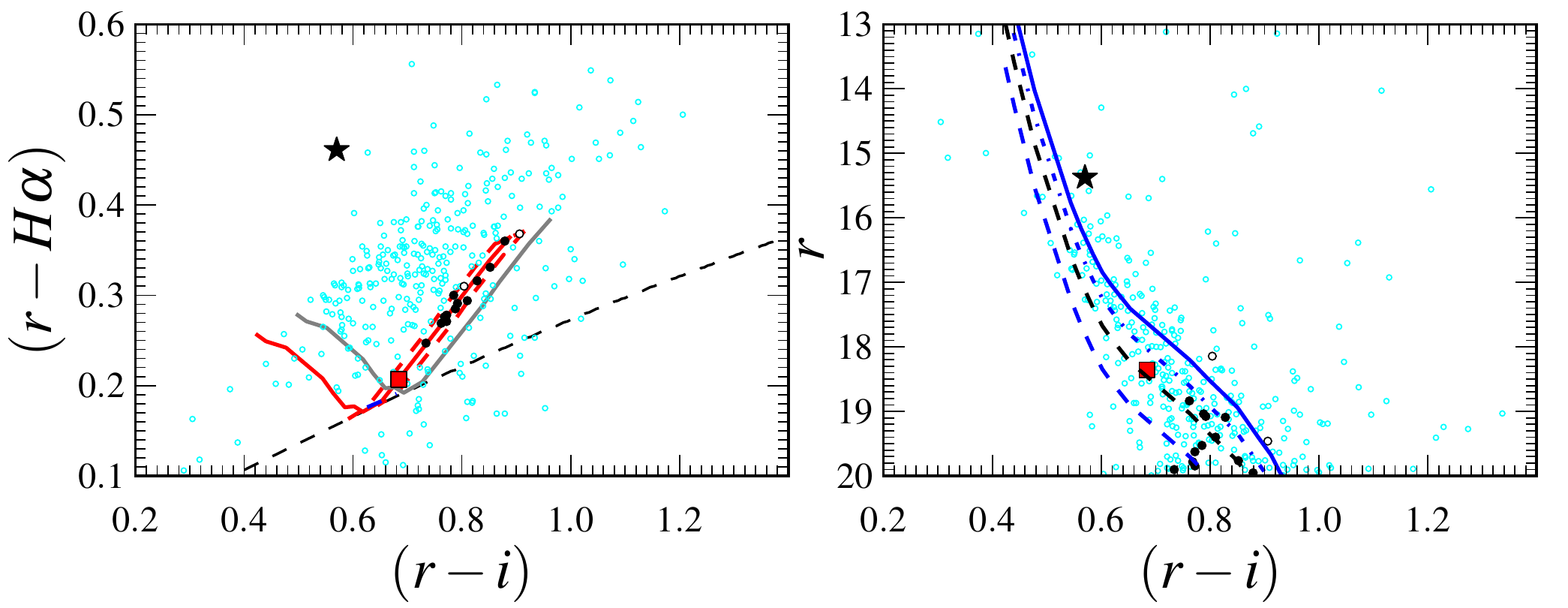}
\label{ms-fit:b}} 
\caption{Two examples of luminosity class assignment based on the IPHAS photometry of stars
selected within a box of $10 \times 10$~arcmin$^{2}$ centred on the
CBe star. The top panel shows the data relevant to 
star \#~11, while the bottom one pertains to star \#~52. To the left,
sightline colour-colour diagrams for stars with $r \leq 20$ are
presented.  The solid red curve in each case is the MS track,
reddened by same amount as the CBe star.  The red-dashed lines are the
tracks for $E(B-V)_{{\rm S,c}} \pm 0.5\sigma$: stars falling between them
are selected as stars of similar reddening -- only those with $(r
-H\alpha)$ colour consistent with their being A--F stars are retained
(these are picked out in black).  The CBe star itself is marked by the
star symbol. The grey MS track also drawn is reddened at the line-of-sight
\citetalias{SFD98} asymptotic colour excess.
The dashed black curve is the early-A reddening line.  The right hand
panels present the colour-magnitude diagrams for the $10 \times
10$~arcmin$^2$ selections. The reddened MS loci, computed for the
distance moduli consistent with the CBe star as (i) class V, (ii)
class IV, or (iii) class III are plotted respectively as solid,
dash-dotted and dashed blue curves. The formal best-fit MS locus
(reddened by the same amount as the CBe star) is
plotted as a dashed black curve. Stars contributing to the fit are the 
black filled dots and the red squares (early A stars), while the
unfilled black circles are stars excluded from the fit.}
\label{ms-fit}
\end{figure*}

\begin{enumerate}

\item The photometry of all the stellar and probably-stellar point sources 
(morphology classification codes -1 and -2) with $r \leq 20$ is collected 
(cyan empty circles), within an on-sky box of $10 \times 10$~arcmin$^{2}$ 
centred on each CBe star (black star in figure~\ref{ms-fit}).

\item The MS track, reddened by the amount corresponding to
  $E(B-V)_{\rm{S,c}}$, is identified (plotted as the red solid curve in the 
left-hand panels of Fig.~\ref{ms-fit}). This was produced by computing 
synthetic $(r,\,i,\,\Halpha)$ photometry from a grid of
\citep[][]{Munari05} 
MS models, that were scaled to the Calspec model Vega spectrum\footnote{Obtained from
http://www.stsci.edu/hst/observatory/cdbs/calspec.html. It is a Kurucz (2005) 
model with T$_{{\rm eff}}$ = 9400~K model with $\log g = 3.90$.}. 

\item All the point sources that fall within the reddening range,
$E(B-V)_{\rm{S,c}} \pm 0.5 \sigma$ 
(dashed red curves), and have the colours appropriate to early-A to 
late-F stars, are selected. The working assumption 
is that these stars, in sharing essentially the same reddening, are likely to 
be as far away as the CBe star.
\setcounter{table}{5}
\begin{table*}
\begin{minipage}{135mm}
\caption[]{Table of spectroscopic parallaxes of the CBe
  stars. Columns list: ID number; spectral type; Galactic coordinates; 
  the observed $r$ magnitude corrected for circumstellar 
  disc emission; $A_{r}$, computed from $E(B-V)_{\rm{(S, c)}}$; A/F fit approximate distances; 
  spectro-photometric distances for luminosity classes V, IV and 
  and III. In bold-face are distances that are associated to the preferred luminosity class, 
  which is noted in the last column.  The error on $D_{\rm{SP}}$ carries independent contributions from: photometric error,
  reddening error, disc emission uncertainty and the spread in absolute magnitude, as given in Table~\ref{intrinsic}. }
\begin{tabular}{@{}rlrrrrrrrrr@{}}
\hline
 & & & & & & \multicolumn{4}{c}{Distances}& \\
\#  & SpT & $\ell$ & $b$ &$r + \Delta r$ &   $A_{r}$ & A/F fit & $D_{\rm{SP, V}}$ & $D_{\rm{SP,IV}}$ & $D_{\rm{SP, III}}$ &  Likely \\
    & &(deg) & (deg) &(mag) &   (mag) & (kpc) & (kpc) & (kpc) & (kpc) &   Class \label{t:distances}\\
\hline
 1 & B5 & 120.04 & 1.64 &$ 14.90 \pm 0.04 $& 3.44 & 5.1  &$ 2.7 \pm 0.3 $&$ 3.4 \pm 0.5 $&$ {\bf 5.6 \pm 1.5} $& III\\
 2 & B7 & 120.45 & 0.32 &$ 14.14 \pm 0.04 $& 1.62 & 2.9  &$ {\bf 3.3 \pm 0.4} $&$ 4.0 \pm 0.6 $&$ 6.6 \pm 1.6 $& V\\
 3 & B3 & 121.09 & 3.99 &$ 14.66 \pm 0.03 $& 3.90 & 4.7  &$ 2.9 \pm 0.4 $&$ 3.6 \pm 0.5 $&$ {\bf 5.2 \pm 1.3} $& III\\
 4 & A0 & 121.40 & 3.92 &$ 16.12 \pm 0.04 $& 2.48 & 3.7  &$ {\bf 3.4 \pm 0.5} $&$ 4.2 \pm 0.7 $&$ 6.7 \pm 2.0 $& V\\
 5 & B2 & 121.76 & 2.43 &$ 15.05 \pm 0.03 $& 2.71 & 3.4  &$ {\bf 7.8 \pm 1.5} $&$ 9.4 \pm 1.6 $&$ 13.5 \pm 3.1 $& V\\
 6 & B3 & 122.26 & 1.16 &$ 15.78 \pm 0.04 $& 3.39 & 7.0  &$ 6.1 \pm 1.0 $&$ {\bf 7.7 \pm 1.1} $&$ 11.1 \pm 2.9 $& IV\\
 7 & B7 & 122.41 & 0.12 &$ 14.98 \pm 0.04 $& 2.05 & 3.0  &$ {\bf 4.0 \pm 0.5} $&$ 4.8 \pm 0.7 $&$ 8.0 \pm 1.9 $& V\\
 8 & B3 & 122.79 & 0.72 &$ 15.54 \pm 0.04 $& 2.71 & 5.5  &$ {\bf 7.5 \pm 1.1} $&$ 9.4 \pm 1.2 $&$ 13.6 \pm 3.3 $& V\\
 9 & B5 & 122.80 & 2.07 &$ 14.26 \pm 0.04 $& 2.30 & 3.1  &$ {\bf 3.4 \pm 0.5} $&$ 4.2 \pm 0.6 $&$ 7.0 \pm 1.9 $& V\\
10 & B7 & 122.83 & 2.69 &$ 14.86 \pm 0.04 $& 2.66 & 4.4  &$ 2.9 \pm 0.4 $&$ {\bf 3.4 \pm 0.5} $&$ 5.7 \pm 1.4 $& IV\\
11 & B2-3 & 123.29 & 0.23 &$ 15.22 \pm 0.04 $& 2.23 & 6.1  &$ {\bf 9.2 \pm 1.5} $&$ 11.3 \pm 1.7 $&$ 16.4 \pm 3.9 $& V\\
12 & B5 & 123.47 & 0.20 &$ 14.50 \pm 0.04 $& 2.10 & 3.4  &$ {\bf 4.1 \pm 0.6} $&$ 5.2 \pm 0.8 $&$ 8.6 \pm 2.4 $& V\\
13 & B5 & 123.49 & 0.11 &$ 14.70 \pm 0.03 $& 1.62 & 6.0  &$ {\bf 5.6 \pm 0.7} $&$ 7.1 \pm 1.1 $&$ 11.8 \pm 3.2 $& V\\
14 & B4 & 123.98 & 0.44 &$ 15.59 \pm 0.04 $& 2.76 & 4.9  &$ {\bf 6.3 \pm 0.8} $&$ 7.9 \pm 1.0 $&$ 11.4 \pm 3.1 $& V\\
15 & B5 & 124.72 & 0.04 &$ 14.71 \pm 0.04 $& 2.68 & 3.8  &$ {\bf 3.5 \pm 0.4} $&$ 4.3 \pm 0.6 $&$ 7.2 \pm 1.9 $& V\\
16 & B3 & 125.04 & 0.08 &$ 14.27 \pm 0.04 $& 3.09 & 7.2  &$ 3.5 \pm 0.5 $&$ 4.4 \pm 0.6 $&$ {\bf 6.3 \pm 1.6} $& III\\
17 & B3 & 125.40 & 3.27 &$ 14.72 \pm 0.04 $& 3.44 & 4.7  &$ 3.6 \pm 0.6 $&$ {\bf 4.6 \pm 0.6} $&$ 6.6 \pm 1.6 $& IV\\
18 & B4 & 126.22 & 1.79 &$ 14.60 \pm 0.04 $& 2.50 & 2.5  &$ {\bf 4.5 \pm 0.6} $&$ 5.6 \pm 0.7 $&$ 8.1 \pm 2.2 $& V\\
19 & B6 & 126.25 & 3.36 &$ 15.13 \pm 0.04 $& 2.78 & 4.1  &$ 3.5 \pm 0.4 $&$ {\bf 4.2 \pm 0.6} $&$ 7.0 \pm 1.9 $& IV\\
20 & B3 & 126.42 & 1.32 &$ 14.45 \pm 0.04 $& 3.24 & 2.1  &$ {\bf 3.5 \pm 0.5} $&$ 4.5 \pm 0.6 $&$ 6.5 \pm 1.6 $& V\\
21 & B5 & 126.47 & 1.24 &$ 15.27 \pm 0.03 $& 3.37 & 3.2  &$ {\bf 3.3 \pm 0.5} $&$ 4.1 \pm 0.7 $&$ 6.8 \pm 1.9 $& V\\
*22 & B4 & 126.55 & 2.64 &$ 14.88 \pm 0.04 $& 3.26 & 9.2  &$ 3.6 \pm 0.4 $&$ 4.5 \pm 0.6 $&$ {\bf 6.5 \pm 1.7} $& III\\
23 & B7 & 126.65 & 3.24 &$ 14.82 \pm 0.03 $& 2.53 & 3.0  &$ {\bf 3.0 \pm 0.4} $&$ 3.6 \pm 0.6 $&$ 5.9 \pm 1.5 $& V\\
24 & B3 & 126.86 & 1.13 &$ 14.19 \pm 0.04 $& 3.31 & 3.0  &$ {\bf 3.0 \pm 0.4} $&$ 3.8 \pm 0.5 $&$ 5.5 \pm 1.4 $& V\\
25 & B5 & 126.87 & -0.10 &$ 13.48 \pm 0.04 $& 2.08 & 2.1  &$ {\bf 2.6 \pm 0.3} $&$ 3.3 \pm 0.5 $&$ 5.4 \pm 1.4 $& V\\
26 & B3 & 127.30 & 2.36 &$ 15.97 \pm 0.04 $& 2.53 & 5.7  &$ {\bf 9.9 \pm 1.6} $&$ 12.4 \pm 1.8 $&$ 18.0 \pm 4.6 $& V\\
27 & B5 & 128.71 & 0.49 &$ 14.65 \pm 0.04 $& 2.66 & 6.5  &$ 3.4 \pm 0.4 $&$ 4.3 \pm 0.6 $&$ {\bf 7.1 \pm 1.9} $& III\\
28 & B7 & 128.74 & 2.55 &$ 14.50 \pm 0.04 $& 2.02 & 3.5  &$ {\bf 3.2 \pm 0.4} $&$ 3.9 \pm 0.6 $&$ 6.5 \pm 1.6 $& V\\
29 & B7 & 128.85 & 1.29 &$ 14.08 \pm 0.04 $& 1.95 & 3.2  &$ 2.8 \pm 0.3 $&$ {\bf 3.3 \pm 0.5} $&$ 5.5 \pm 1.4 $& IV\\
30 & B4 & 129.19 & 2.60 &$ 14.63 \pm 0.04 $& 1.95 & 4.9  &$ {\bf 5.8 \pm 0.7} $&$ 7.3 \pm 0.9 $&$ 10.6 \pm 2.8 $& V\\
31 & B3 & 129.46 & 0.68 &$ 13.85 \pm 0.04 $& 3.14 & 3.9  &$ 2.8 \pm 0.4 $&$ {\bf 3.5 \pm 0.4} $&$ 5.1 \pm 1.2 $& IV\\
32 & B6 & 129.81 & 3.87 &$ 15.31 \pm 0.04 $& 2.45 & 5.2  &$ 4.4 \pm 0.6 $&$ {\bf 5.3 \pm 0.8} $&$ 8.8 \pm 2.4 $& IV\\
33 & B8-9 & 129.82 & 1.00 &$ 14.45 \pm 0.03 $& 2.18 & 4.9  &$ 2.4 \pm 0.3 $&$ 3.0 \pm 0.5 $&$ {\bf 4.7 \pm 1.2} $& III\\
34 & B3 & 129.97 & 1.96 &$ 14.26 \pm 0.03 $& 1.62 & 3.5  &$ {\bf 6.8 \pm 1.0} $&$ 8.6 \pm 1.1 $&$ 12.4 \pm 3.1 $& V\\
35 & B2-3 & 130.30 & 2.08 &$ 15.40 \pm 0.04 $& 2.12 & 5.2  &$ {\bf 10.4 \pm 1.8} $&$ 12.8 \pm 1.9 $&$ 18.6 \pm 4.4 $& V\\
36 & B4 & 130.41 & -0.59 &$ 14.65 \pm 0.03 $& 3.62 & 2.9  &$ {\bf 2.7 \pm 0.4} $&$ 3.4 \pm 0.5 $&$ 5.0 \pm 1.4 $& V\\
37 & B6 & 131.56 & 1.01 &$ 14.53 \pm 0.04 $& 2.25 & 3.2  &$ {\bf 3.4 \pm 0.5} $&$ 4.1 \pm 0.6 $&$ 6.7 \pm 1.8 $& V\\
38 & B5 & 131.92 & 1.32 &$ 15.66 \pm 0.04 $& 2.20 & 3.7  &$ {\bf 6.7 \pm 1.3} $&$ 8.4 \pm 1.8 $&$ 14.0 \pm 4.3 $& V\\
39 & B4 & 132.86 & 1.81 &$ 16.25 \pm 0.04 $& 2.08 & 4.5  &$ {\bf 11.6 \pm 2.0} $&$ 14.6 \pm 2.5 $&$ 21.1 \pm 6.1 $& V\\
*40 & B2 & 133.79 & 2.35 &$ 14.38 \pm 0.03 $& 2.58 & 11.9  &$ 6.0 \pm 1.1 $&$ 7.3 \pm 1.2 $&$ {\bf 10.5 \pm 2.4} $& III\\
41 & B7 & 134.13 & -0.59 &$ 14.11 \pm 0.04 $& 2.50 & 2.5  &$ 2.2 \pm 0.3 $&$ {\bf 2.6 \pm 0.4} $&$ 4.3 \pm 1.1 $& IV\\
42 & B2 & 134.49 & -0.49 &$ 14.95 \pm 0.03 $& 3.44 & 3.4  &$ {\bf 5.3 \pm 1.0} $&$ 6.4 \pm 1.1 $&$ 9.2 \pm 2.1 $& V\\
44 & B5 & 135.03 & 1.41 &$ 15.50 \pm 0.04 $& 1.97 & 3.2  &$ {\bf 6.9 \pm 0.9} $&$ 8.7 \pm 1.2 $&$ 14.4 \pm 3.8 $& V\\
45 & B3 & 135.06 & 0.55 &$ 13.01 \pm 0.04 $& 2.81 & 2.6  &$ 2.2 \pm 0.3 $&$ {\bf 2.8 \pm 0.3} $&$ 4.0 \pm 1.0 $& IV\\
46 & B9 & 135.14 & -0.81 &$ 14.61 \pm 0.04 $& 2.05 & 2.4  &$ {\bf 2.5 \pm 0.3} $&$ 3.1 \pm 0.5 $&$ 4.9 \pm 1.4 $& V\\
47 & B3 & 135.38 & -0.08 &$ 13.87 \pm 0.03 $& 4.00 & 2.6  &$ 1.9 \pm 0.3 $&$ {\bf 2.4 \pm 0.3} $&$ 3.5 \pm 0.8 $& IV\\
49 & A0 & 135.64 & 2.19 &$ 15.42 \pm 0.04 $& 1.77 & 6.1  &$ 3.4 \pm 0.5 $&$ 4.3 \pm 0.7 $&$ {\bf 6.8 \pm 1.9} $& III\\
50 & B3 & 135.89 & 1.30 &$ 16.03 \pm 0.04 $& 2.33 & 5.5  &$ {\bf 11.1 \pm 1.7} $&$ 14.0 \pm 1.9 $&$ 20.2 \pm 5.0 $& V\\
51 & B8-9 & 136.07 & 2.94 &$ 14.61 \pm 0.04 $& 1.80 & 8.7  &$ 3.0 \pm 0.4 $&$ 3.8 \pm 0.6 $&$ {\bf 6.0 \pm 1.6} $& III\\
52 & B7 & 136.09 & 1.48 &$ 15.49 \pm 0.04 $& 2.20 & 6.9  &$ 4.7 \pm 0.6 $&$ {\bf 5.7 \pm 0.8} $&$ 9.4 \pm 2.3 $& IV\\
53 & B7 & 136.14 & 0.42 &$ 14.33 \pm 0.03 $& 2.63 & 2.1  &$ {\bf 2.3 \pm 0.3} $&$ 2.7 \pm 0.4 $&$ 4.5 \pm 1.1 $& V\\
54 & B3 & 136.15 & 1.70 &$ 15.75 \pm 0.04 $& 2.05 & 5.2  &$ {\bf 11.1 \pm 1.6} $&$ 14.0 \pm 1.7 $&$ 20.3 \pm 5.0 $& V\\
55 & B3-4 & 136.17 & 1.32 &$ 16.20 \pm 0.04 $& 2.48 & 6.8  &$ {\bf 10.3 \pm 1.4} $&$ 13.0 \pm 1.7 $&$ 18.7 \pm 4.9 $& V\\
56 & B7 & 136.27 & 0.59 &$ 13.99 \pm 0.04 $& 2.38 & 2.4  &$ {\bf 2.2 \pm 0.3} $&$ 2.6 \pm 0.4 $&$ 4.3 \pm 1.1 $& V\\
58 & B3 & 136.34 & 1.90 &$ 14.10 \pm 0.04 $& 1.70 & 5.2  &$ {\bf 6.1 \pm 0.9} $&$ 7.7 \pm 1.0 $&$ 11.1 \pm 2.7 $& V\\
\hline
\multicolumn{11}{p{\linewidth}}{Note: $^{*}$ For these sightlines the
  A/F fit distance estimate is based on 2 or 3 nearby early-A stars
  alone.}
\end{tabular}
\end{minipage}
\end{table*}
\begin{table*}
\begin{minipage}{135mm}
\contcaption{}
\begin{tabular}{@{}rlrrrrrrrrr@{}}
\hline
 & & & & & & \multicolumn{4}{c}{Distances}& \\
\#    & SpT & $\ell$ & $b$ &$r + \Delta r$ &   $A_{r}$ & A/F fit & $D_{\rm{SP, V}}$ & $D_{\rm{SP,IV}}$ &$D_{\rm{SP, III}}$ &  Likely \\
    & &(deg) & (deg) &(mag) &   (mag) & (kpc) & (kpc) & (kpc) & (kpc) &   Class \\
\hline
59 & B5 & 136.50 & 2.28 &$ 14.41 \pm 0.03 $& 1.77 & 5.8  &$ 4.6 \pm 0.6 $&$ {\bf 5.8 \pm 0.9} $&$ 9.6 \pm 2.6 $& IV\\
60 & B3-4 & 136.50 & 2.26 &$ 15.50 \pm 0.04 $& 1.95 & 6.3  &$ {\bf 9.5 \pm 1.3} $&$ 12.0 \pm 1.6 $&$ 17.3 \pm 4.5 $& V\\
61 & B7 & 136.64 & 2.38 &$ 14.89 \pm 0.04 $& 1.82 & 4.3  &$ {\bf 4.3 \pm 0.6} $&$ 5.1 \pm 0.8 $&$ 8.5 \pm 2.1 $& V\\
62 & B6 & 137.34 & 1.60 &$ 16.24 \pm 0.04 $& 3.04 & 2.7  &$ {\bf 5.2 \pm 0.8} $&$ 6.2 \pm 1.0 $&$ 10.3 \pm 2.9 $& V\\
64 & B5 & 138.63 & -0.27 &$ 16.33 \pm 0.04 $& 3.09 & 3.1  &$ {\bf 6.1 \pm 0.8} $&$ 7.6 \pm 1.2 $&$ 12.7 \pm 3.5 $& V\\
65 & B5 & 138.81 & -0.87 &$ 13.97 \pm 0.04 $& 2.35 & 1.8  &$ {\bf 2.9 \pm 0.4} $&$ 3.6 \pm 0.5 $&$ 6.0 \pm 1.6 $& V\\
66 & B4 & 138.98 & -0.94 &$ 14.40 \pm 0.04 $& 2.68 & 2.8  &$ {\bf 3.7 \pm 0.5} $&$ 4.7 \pm 0.6 $&$ 6.8 \pm 1.8 $& V\\
67 & B4 & 139.21 & 2.58 &$ 15.22 \pm 0.04 $& 2.81 & 3.6  &$ {\bf 5.2 \pm 0.8} $&$ 6.5 \pm 1.0 $&$ 9.4 \pm 2.7 $& V\\
\hline
\end{tabular}
\end{minipage}
\end{table*}

\item We estimate the distance to the group of stars selected from the
  IPHAS colour-magnitude diagram by finding the MS track that fits
  them best (dashed black curve in each of the right-hand panels).
In the fitting procedure, the selected stars are weighted according to
their photometric errors and with a sigmoid function computed as 
described by \citep[][]{Sale09}.  The latter limits the bias to too 
short a distance that is otherwise induced by stars just brighter than 
the magnitude limit. Furthermore, since the IPHAS colours roughly
signal spectral type, and early-A candidates are the least ambiguous,
extra weight was awarded to them (5 times that of other later-type stars).
We also applied a 3-$\sigma$ cut to bright (redder) outliers 
and recomputed the fit, in order to inhibit shortening of the distance due 
to interloping giant stars. The MS absolute magnitude scale applied to
the selected A and F stars is taken from \citet{Sale09} 
The distances estimated this way are reported in
Table~\ref{t:distances}.  In view of the modest samples sizes
involved, these distances are indicative only and certainly
approximate, and used here solely as a guide to likely luminosity class.

\item A luminosity class (either V, IV or III, given in the final column of
Table~\ref{t:distances}) is then assigned to each CBe star 
according to the option falling closest to the rough distance estimate from 
MS-fitting. In Fig.~\ref{ms-fit}, the MS loci consistent with class V,
IV and III luminosity-class assignments for the CBe star are plotted in the 
colour-magnitude diagrams.  Where the distance estimate obtained from
the candidate A--F stars is lower than the class V spectroscopic parallax, 
$D_{\rm{S,\,V}}$, the (longer) distance compatible with class V is
adopted. 
\end{enumerate}
\setcounter{figure}{10}
\begin{figure*}
\centering
\includegraphics[width=\linewidth]{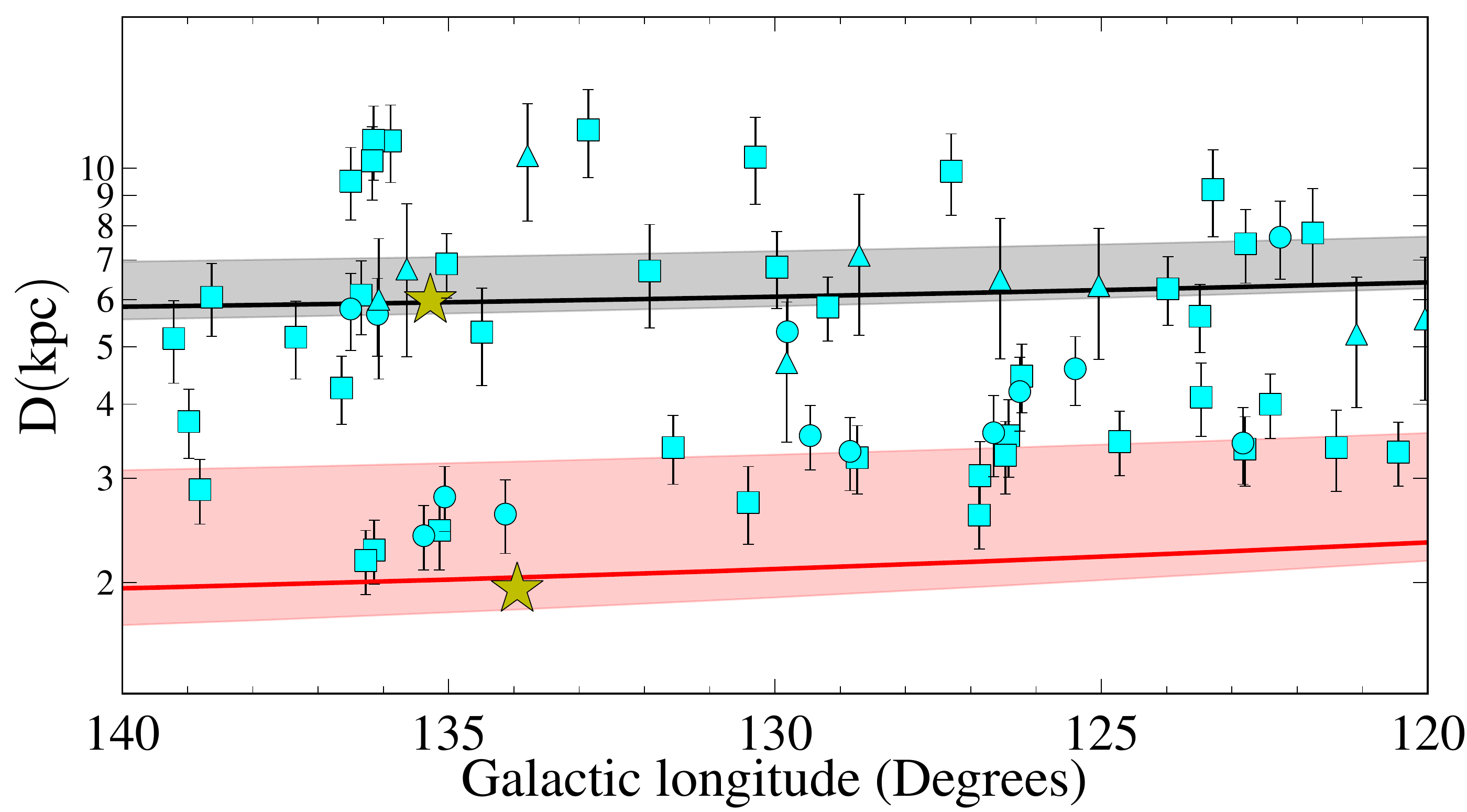}
\caption{The spatial distribution of our CBe stars on the Galactic
  Plane is shown. Different symbols are used for the luminosity class: 
  dwarf (squares), circles (sub-giants), triangles (giants).
  The spiral arms are plotted following the prescriptions given in
  \citet{Vallee08} -- solid red curve for the Perseus Arm, black solid curve for the Outer Arm. 
  Instead, the two bands of width 1.4~kpc \citep{Russeil07}, represent the range of distances 
  that are covered by the two spiral arms in \citep{Russeil07}. 
  The Perseus Arm is in pink, the Outer Arm is in grey. The two yellow
  stars mark the trigonometric parallaxes of masers as specified by 
  \citet{Reid09}, which are sitting on the near edge of the spiral arms where the star formation
  is active.}
\label{arms_distances_phot}
\end{figure*}

 As a partial test of this method of estimation, we have applied
  it to photometric selections of A/F stars
in the well-studied clusters, NGC~637 and NGC~663. In the case
of NGC~637, \cite{Yadav08} obtained $2.5 \pm 0.2$~kpc from
conventional photometric methods -- our shorthand method gives
2.0~kpc. For NGC~663, we obtain A/F distances for three different sightlines
crossing the cluster (to CBe stars for which we have only FLWO-1.5m/FAST
spectra) that are respectively 2.2, 2.5, and 2.8~kpc.  These compare 
satisfactorily with the literature measure of $2.4 \pm 0.1$~kpc 
\citep{Pandey05} for this cluster.  Nevertheless the method carries
two potential biases towards low estimates that results in 16
A/F star distances that are much lower than the class V CBe-star 
distance.  First, it rests on trying to identify associated
main-sequence A/F stars -- although we attempt to eliminate
interloping giants (improbable companions for CBe stars), this may not 
always be successful. Second, where the CBe star is relatively
early-type (B2-4) and very distant, the reddening may be comparable
with the total Galactic reddening with the result that the detected 
A/F stars may actually be foreground and unassociated.  The first of 
these biases may result in inappropriate assignment to class V,
but the second most likely 'fails safe' (in 9 of the 16 cases) in
leaving these objects as dwarfs at distances of between 8 and 12 kpc. 
Given these issues, we do not regard these estimates as providing more 
than an ad hoc sorting tool.

The pattern emerging from the luminosity class assignments is similar
to that among the sample of classical Be stars presented by \citet{ZB97}:
42 are assigned to class V (cf 36,  on scaling to this older result), 
while 12 and 9 are placed into classes IV and III respectively (cf 
expectations of 14--15, and 13).  That there are more dwarfs may 
 either be a consequence of the much fainter apparent magnitude
  range our sample is drawn from, or due to the noted bias in the
  method of assignment.

\subsection{Spatial distribution of the CBe sample}
\label{chap4.3}
In Fig.~\ref{arms_distances_phot} we plot all the stars at the distance 
corresponding to their assigned luminosity class against Galactic longitude,
marking on the diagram the expected locations of the Perseus and Outer 
Arms. The four stars for which we do not have spectroscopic $\Halpha$ 
observations are not included in this plot. The emergent picture presented
by these 63 stars is certainly not one of pronounced clustering
picking out the spiral arms in the distance-longitude plot.  Closest
to this possible reality is seen at longitude $\sim$135 where there is
a group of six stars near the star-forming complexes W3/W4/W5, 
well in front of another group of stars, sitting closer to the OH maser in 
the Outer Arm.  Elsewhere there is no sign of such orderly behaviour.
The casual impression is of a scattered, more or less random,
distribution of emission line stars.

In the sample, no CBe star is closer than $2.2\pm
0.3$~kpc (\#~56) or more distant than $11.6\pm 2.0$~kpc (\#~39).
This is mainly a reflection of the magnitude limits ($13 \lesssim r
\lesssim 16$) 
placed on the sample of CBe stars.  At the bright end $(r = 13)$, a
main sequence dwarf with a median spectral type (B5),
with a median reddening of $A_r = 2.5$, just falls within the sample at
the minimum distance of $\sim 2.0$~kpc.  For B3V this estimate of the minimum 
rises to 2.9~kpc, consistent with the brightest object (\# 45) in the
sample, that happens to be a B3 star, being assigned a distance of
2.8~kpc (its reddening is a little above the median value).  For the 
latest spectral types present in the sample, the near distance limit
drops as low as 1 to 1.5~kpc.  That we do not find any in the allowed
range between 1 and 2 kpc is perhaps because the reddening only rises
up to and through the median for the distribution once the Perseus Arm
is well and truly entered at $\sim$2~kpc.  

The upper distance limit can in principle be expected to be more
variable, following to an extent the variation of the integrated
Galactic reddening with sightline \citepalias[][]{SFD98}: for most of the CBe sample 
the maximum possible $A_r$ varies from $\sim$2 up to $\sim$5.  But
our selection has, for practical observational reasons, avoided the
most heavily reddened objects and sightlines (the maximum $A_r$ in the
sample is 4).  On deploying the median spectral type and
reddening for the sample, again -- but this time combining them with
the absolute magnitude appropriate to luminosity class III --- we
would expect a faint magnitude limit of $r \sim 16$ to translate to a 
maximum heliocentric distance of $\sim$16~kpc (dropping to 10--12~kpc
for the latest B sub-types).  The actual outcome is that the most 
distant/faintest objects in the sample are B3-4Ve objects inferred to 
be 10--12~kpc away.  The objects assigned to luminosity class III are,
in contrast, mostly relatively bright and/or relatively heavily
reddened, bringing all but one of them in to distances closer than
10~kpc.  So whilst there is not a simple upper limiting distance to
the observed window, there is reason to assume that the range from 3
to 8~kpc is well captured by our sample at all sub-types -- so if CBe
stars are preferentially located in the Outer Arm at 5 to 6~kpc, it
would be likely to be evident.  Fig.~\ref {arms_distances_phot} does
not support this.  We return to this below in the discussion.

The most distant early-type dwarf stars at heliocentric distances of
10--11~kpc are 16--17~kpc away from the Galactic Centre. This places
them significantly outside the disc 'truncation' radius estimated by 
\citet{Ruphy96},  and since re-examined by \citet{Sale10}.  Although 
dependent in detail on how the stellar density profile steepens at 
these large Galactocentric radii, we would not expect a selection of CBe
stars fainter than $r = 16$ to yield too many more-distant objects -- 
instead, it would more likely add in stars that are later in spectral type, 
more reddened, or both.  Indeed the number of early-type stars  that
are already known at such a large Galactocentric radii is very small. 
In \citet{Rolleston00}, just 14 out of the 80 studied B-type stars, 
between $6 \leq R_{G} \leq 18$~kpc,
are found at distances larger than $R_{G} \sim 13$~kpc.
\section{Discussion}
\label{chap5}
In this section, we identify the main insights provided by our sample
of 67 CBe stars, identifying robust outcomes and possible biases. 
Regarding the latter, we analyse the impact that choices of reddening 
law and absolute magnitude scale, and the method of correction for 
circumstellar disc fraction, may have had on the distance estimates.
Finally, we compare (i) the inferred cumulative distribution of object 
distances with that expected of a regularly declining disc stellar density 
profile, (ii) the measured corrected colour excesses,
$E(B-V)_{\rm{S,\,c}}$, with the integrated colour excesses from 
\citetalias{SFD98}.  
\subsection{Possible biases}
We now turn to the individual biases that may affect the distances
inferred for our sample.

\subsubsection{The absolute magnitude scale and luminosity classification}
\label{chap5.1.1}
In Section~\ref{chap4.1}, we pointed out that our chosen absolute magnitude 
scale is the faintest among those to be found in current literature. 
For example the MS magnitudes we have adopted are, on average,
0.4-0.6~mag fainter than others reported in literature 
\citep[see e.g.][]{Straizys81, Aller82, Wegner00} for the early and 
late-B types,
whilst they are better aligned for mid-B stars. Had we favoured a 
brighter absolute magnitude scale, we should expect to obtain
distances up to 25~\% larger than those we have tabulated.  However
it is worth noting that we found that the great majority of our
class V spectroscopic parallaxes gave larger values than those crudely
inferred from nearby candidate A/F stars (section~\ref{chap4.2} and 
Table~\ref{t:distances}).  This may turn out to be part of the
explanation for the attribution of a somewhat higher proportion of 
the sample to class V, based on the existing absolute magnitude scale,
relative to the earlier sample of \citet{ZB97}. 

The deduced distance to each CBe star is necessarily strongly
dependent on adopted luminosity class.  Here, a rough constraint on 
luminosity class has come from estimating the distances to
probable main-sequence A and F stars of comparable reddening within
a few arcminutes angular separation (section~\ref{chap4.2}).  
Where the luminosity assignment is wrong by one class, 
the distance will be over or under-estimated by 30 \% -- a large
uncertainty. To do better, a robust spectroscopic indicator is needed.
One possibility under investigation presently is to adapt the
Barbier-Chalonge-Divan (BCD) method that evaluates the properties
of the Balmer limit (Fabregat et al, in preparation) to the
circumstances of the entire FLWO-1.5m/FAST dataset. This will 
incorporate the subset of stars that we have described here in detail.  
For the time being it is reassuring that the spread of this smaller
sample across luminosity class is not radically different from that
found by \citet{ZB97}.

At the present time, the uncertainty in absolute magnitude is the major
contribution to the error budget for the distance determinations. The errors
are large enough that attention must be paid to an error-sensitive statistical
bias that was discussed in \citet{Feast72}. This is dependent on the gradient 
in space density and its impact is quantified in Section~\ref{chap5.2} where 
we discuss the spatial distribution of the sample.

\subsubsection{Disc fraction estimates}
\label{chap5.1.2}
We based our disc fraction and circumstellar excess estimates using the 
commonly adopted method proposed by \citet{Dachs88}, in which 
the measured $E(B-V)_{\rm{S}}$ is corrected downwards using a scaling
to the $\Halpha$ emission equivalent width. As we already noticed in 
Sec.\ref{chap3.3}, we observed 5 stars with $EW(\Halpha) \leq -60$~\AA, 
that lie outside the range in which Eq.\ref{dachs:a} and \ref{dachs:b} were
defined. This makes their $E^{cs}(B-V)$ determination more uncertain. 
Furthermore, both of \citep{Dachs88} equations are based on quite
scattered data. 

For the most extreme emitter in our sample (\#~26, with $EW(\Halpha)
\approx -100$~\AA ), a  variation of $f_{D}$ twice as large as the 0.02 
uncertainty that we considered in the error propagation would move the
star by $\pm 5$\% around its measured
distance, after taking into account the corresponding $E^{cs}(B-V)$ and
$\Delta r$ changes. The estimate of $E^{cs}(B-V)$ also has an effect
on the identification of the preferred luminosity class, in that
a different $E(B-V)_{\rm{(S,\, c)}}$ value alters the selection of
stars in the colour-colour diagrams of Sec.\ref{chap4.2} and, hence, 
the A/F star fits.  In short, the role of the disc fraction and the
uncertainties in its estimation is complex.  It is fortunate that
for most of the sample its impact is not likely to be very large

\subsubsection{Choice of reddening law}
\label{chap5.1.3}
As we mentioned in Section~\ref{chap3.2} a different choice of $R_V$ would
affect the distance estimates, although the measured colour excesses would 
not change too much since the shape of the curve does not change
significantly if $R_V$ is altered by a few tenths. A smaller/larger
$R_V$ produces lower/higher reddening for a given colour excess, and
hence a larger/smaller inferred spectroscopic parallax.  A study of
the shape of reddening laws across much of the Galactic 
Plane was undertaken by \citet{FitzMassa07}.  Taken at face value, this
work would seem to imply a lower $R_V$ of $2.9 \pm 0.2$ within the
region delimited by $\ell = (120^{\circ}, 140^{\circ})$ and
 $b = (-5^{\circ}, +5^{\circ})$. However this is based only on three
bright B stars, that apparently lie on the near side of the Perseus
Arm. Since the majority of our stars are appreciably more distant than
2~kpc, there is no strong incentive yet to stray from the widely accepted
mean law ($R_V = 3.1$).  Had we preferred $R_V = 2.9$, the derived 
spectroscopic parallaxes would be about $8 - 10 \%$ larger than
specified here.  Conversely, raising $R_V$ above the typical Galactic
value would shorten the distance scale. If a change in either sense is
necessary, it is more likely that $R_{V}$ should be increased.

\subsection{The cumulative distribution of CBe-star distances}
\label{chap5.2}

In this part of the Galactic Plane ($120^{\circ} \leq \ell \leq 140^{\circ}$
and $-1^{\circ} \leq b \leq 4^{\circ}$), Galactic models
\citep{Russeil03, Vallee08} place the Perseus Arm at $\sim 2$~kpc
and the Outer Arm at $\sim 6$~kpc, consistent with measured maser 
parallaxes \citep{Reid09}.   We showed in
Fig.~\ref{arms_distances_phot} the distances to 63 of
the 67 objects presented in this paper as a function of Galactic
longitude, leaving out 4 objects for which we do not have the
H$\alpha$ emission equivalent width data needed to correct the
measured reddenings for circumstellar emission, and one further star
that is more likely to be a YSO. Neither this figure, nor 
the binned histogram distribution shown in Fig.~\ref{histogram_distances:a},
displays a pronounced clustering consistent with these mooted spiral
arm locations.

 We reconsider the distribution collapsed into a cumulative form that
permits an analysis free of binning effects
(Fig.~\ref{histogram_distances:b}).   The magenta curve shown in it is
the cumulative distribution as a function of distance obtained when
all CBe stars are classified as dwarfs, while the blue curve is 
the result obtained on assigning luminosity classes as given in 
Table~\ref{t:distances}.   If the CBe stars were preferentially
located in the Perseus and Outer Arms, we might expect to see
steepenings of the cumulative distribution curve (CDC) in the distance 
ranges associated with the Arms (picked out in the figure).

\setcounter{figure}{11}
\begin{figure}
\centering
\subfigure[]{
\includegraphics[width=\linewidth]{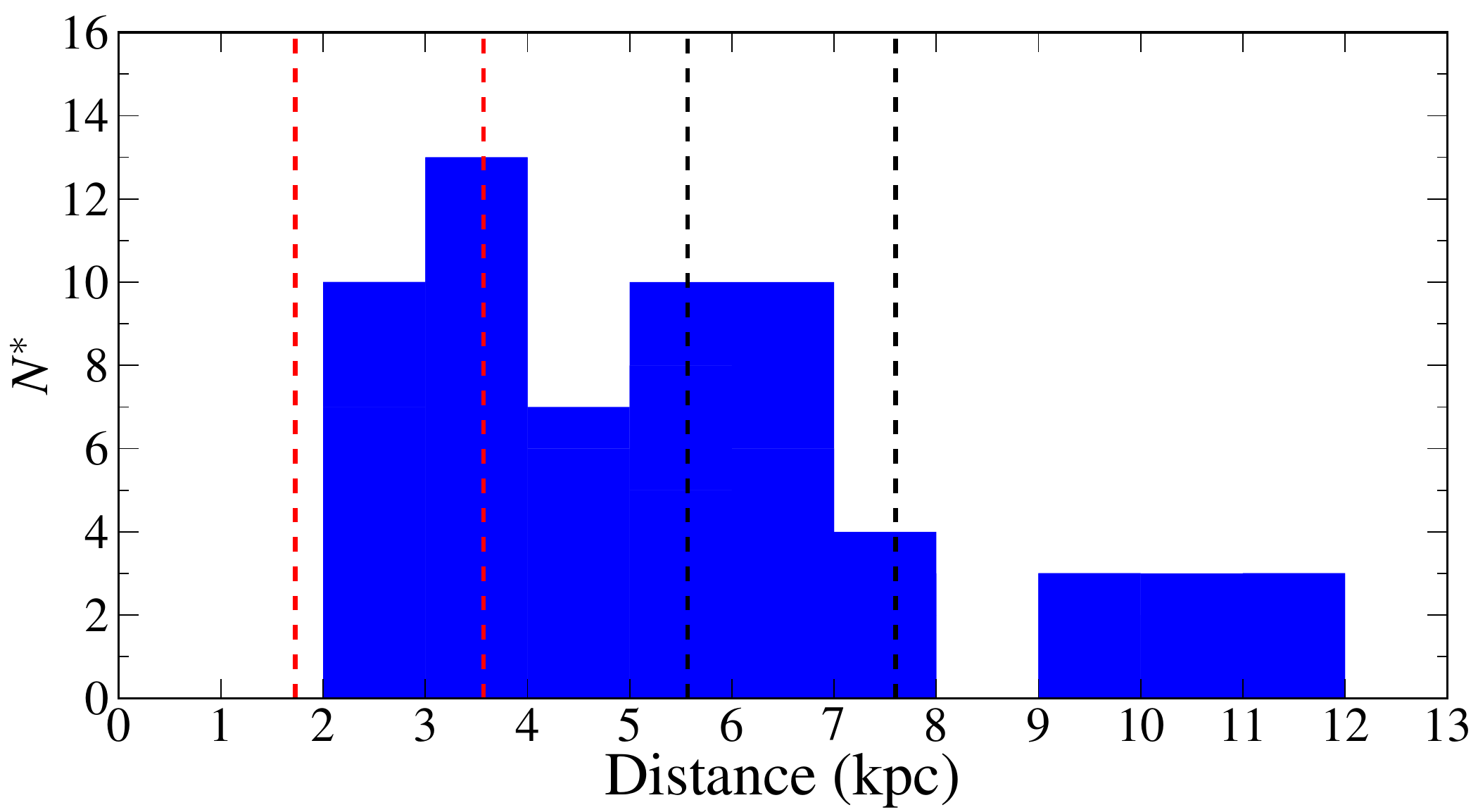}
\label{histogram_distances:a}}
\subfigure[]{
\includegraphics[width=\linewidth]{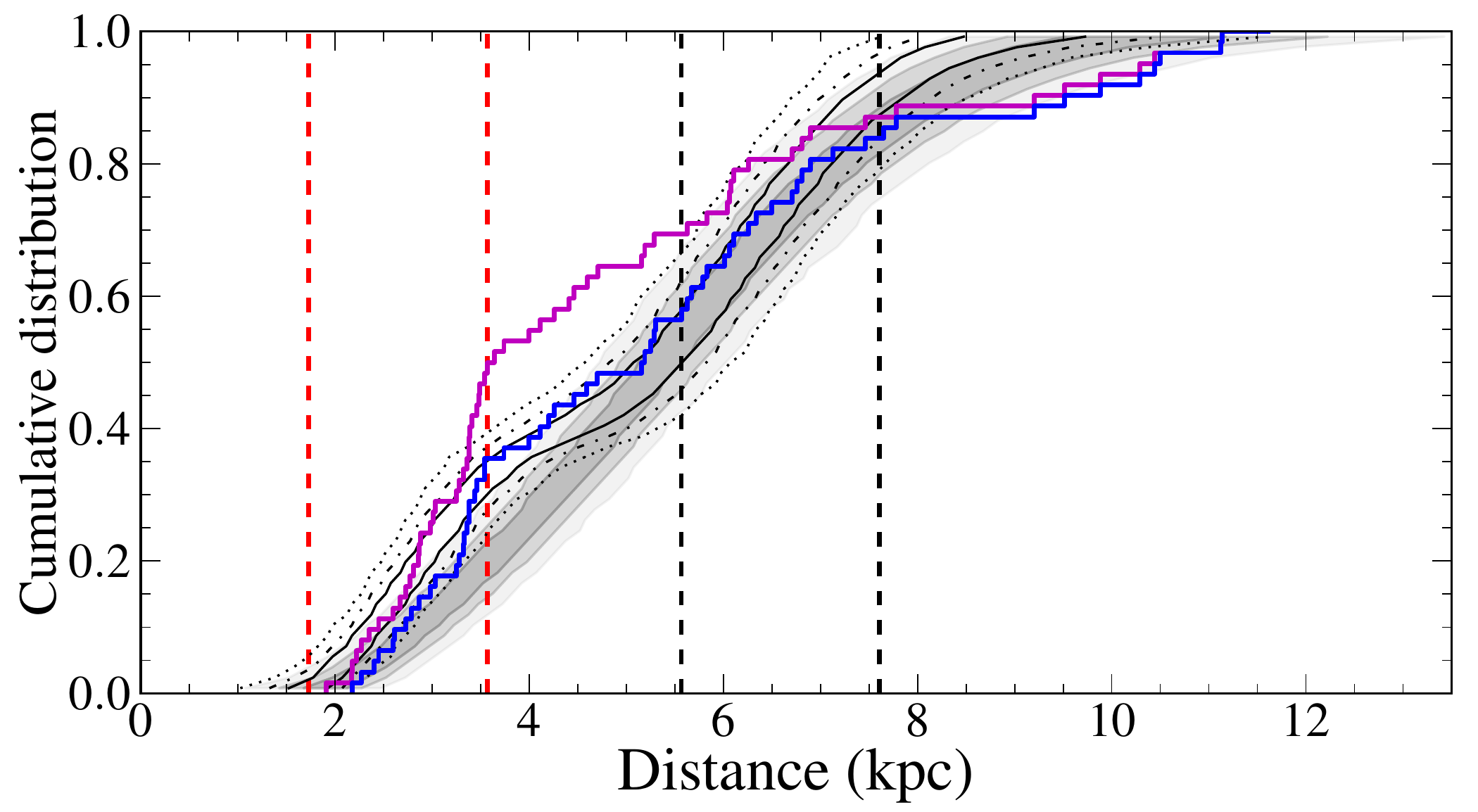}
\label{histogram_distances:b}} 
\caption{(a) Histogram distribution of stars with luminosity class being assigned accordingly
to Table~\ref{t:distances}. The vertical dashed lines define the distance ranges 
thought to be occupied by the spiral arms: red is used for the Perseus
Arm, black for the Outer Arm. (b) The cumulative distribution of CBe stars is plotted against the 
distances derived from spectroscopic parallax. The blue curve expresses the distribution of stars that takes
note of the preferred luminosity class assignment in Table~\ref{t:distances}.  
The dark magenta curve is the result obtained if all the CBe stars are assumed 
to be class V.  The grey-scale contours represent the 1$\sigma$, 2$\sigma$, and 3$\sigma$
confidence limits that result from the MC simulations of a smooth distribution of stars. Similarly, 
the black curves are for the boundaries defined by the family of CDCs, which are derived from the 
MC simulations of stars within the spiral arms. Vertical dashed lines define the range 
of distances associated to the spiral arms, as in panel (a).}
\label{histogram_distances}
\end{figure}

 To test this expectation, we compare our result with simulated, 
appropriately randomised CDCs computed using two contrasting
models: (i) a stellar density gradient consistent with the average
properties of the outer Galactic disc; (ii) a simple spiral arm model
in which it is assumed the CBe stars are contained within them. To
obtain such CDCs, in the first case we set up a distribution function
to be obeyed by the 63 stars that deploys the length scales and
disc 'truncation radius' derived by \citet{Sale10}: essentially the 
exponential length scale out to $R_G = 13 \pm 0.5$~kpc is $(3.0 \pm
0.12)$ kpc, and thereafter it shortens to $(1.2 \pm 0.3)$~kpc.  In 
the second case, we distribute the stars along the line-of-sight
according to two boxcar functions, whose limits are defined by 
the allowed range of distances for the Perseus and Outer Arms given in 
\citet{Russeil07} -- the relative weight of the two spiral arms is set to 
match the exponential decay of case (i). Both distributions are weighted 
with a $D^{2}$
term, to reproduce the conical volume sample function. To emulate the
effect of error in the real data, the randomly selected distances of
stars in each simulation are scattered according to gaussian noise,
that is modelled as a linear function of distance, fit to the real
errors. 10000 Monte Carlo (MC) simulations were performed for each 
type of model. The starting distance of the two models, set to roughly
match the observational selection, does influence the outcome.  
But we find that placing it anywhere between 1 and 2~kpc does not affect 
the median CDC produced by the MC simulations. 
Because of the steep decline in stellar density outside the truncation 
radius, the end point is not influential.  

 We plot both comparison CDCs in Fig.~\ref{histogram_distances:b}, in the form of
contours defining the 1$\sigma$, 2$\sigma$, and 3$\sigma$ confidence limits 
derived from the two families of MC simulations.
A direct visual comparison between the CDC of the 63 CBe stars (blue curve) 
and the contours generated with the simulations, indicates that the
observations incorporating luminosity class constraints (blue curve)
do not clearly prefer either model yet.  However, the CDC 
obtained with all the stars classified as dwarfs (magenta curve) is
exposed as implausible, since too many stars are assigned to the
Perseus Arm.  This is on top of the improbability that all objects in the
CBe sample would be dwarfs, given the known properties of these stars.

 We have performed K-S tests comparing the observed cumulative
  functions with the median of the simulated data, from both models.
For the magenta curve (all stars being dwarfs), we obtain $D_{\rm{no-arms}} = 0.3$ and
$p_{\rm{no-arms}} \sim 0$, $D_{\rm{arms}} = 0.21$ and
$p_{\rm{arms}} = 0.12$; due to the large D values and the small p-values, 
we can reject the hypothesis that the magenta distribution is consistent with either 
models.  On the other hand, for the blue curve we measure $D_{\rm{no-arms}} = 
0.16$ and $p_{\rm{no-arms}} = 0.37$, $D_{\rm{arms}} = 0.14$ and
$p_{\rm{arms}} = 0.51$. In this case the numerical outcome is
inconclusive, rather than negative -- our CBe sample may be compatible
with either model, and it is clear that reduced errors, combined 
perhaps with a larger sample would be needed for a more decisive outcome.

 Furthermore, the astrophysical point that $\sim 60$\% of the stars in  
our sample are B5 or later in spectral type should not be overlooked:
if the Be phenomenon is due to evolutionary structural changes 
\citep{Fabregat00}, these later-type stars would be less likely to
have remained within the spiral arms at an age approaching 
50~Myr or more, than their earlier-type cousins. Ideally, a larger
sample restricted to early-type Be stars, if feasible, would supply
the best test for the spiral arm structure of the Galaxy.

We note that the statistical bias of the type
first set out by \citet{Feast72} and implemented by \citet{Balona74} is present here.  
Since the error model adopted for the MC simulations is based on the uncertainties
affecting the real data, the impact of the bias can be gauged numerically just
by comparing the error-free model CDC with its form when the error is included.
We find, in accordance with expectation, that the shorter distances in the sample are
under-estimated, and the longest are over-estimated.  The effect is most
severe at the longest distances ($> 8$~kpc) where the over-estimation may
approach 1 kpc.  At $\sim 2$~kpc, the under-estimation is in the region of
100-200 pc. Since our MC simulation already takes into account this bias,
the outcome of the K-S test reported above also accounts for it.

Before now, by means of OB-star spectroscopic 
parallaxes derived from a brighter sample ($8 < V < 13$) of stars than here, 
\citet[][]{Negu03} identified a number of OB stars at $d \approx 5.3$~kpc 
for $115^{\circ} < \ell < 120^{\circ}$ and $d = 5$~--~6~kpc for 
$\ell = 175^{\circ}$~--~$215^{\circ}$. Thanks to the high quality of their 
spectroscopic data, the authors felt able to claim spectral
types had been determined to a precision better than $\pm 1$ subtype, 
yielding distances with errors less than 10\%. They surmised that these
objects could belong to the Outer Arm, but stopped short of claiming 
detection of a spiral arm, as such.  Within the same Galactic longitude range 
as considered here, they had very few stars at their disposal.  Here
we have filled in this gap -- but still we do not claim that either
Fig.~\ref{arms_distances_phot} or the preferred blue curve 
in Fig.~\ref{histogram_distances:b}, accounting for the range of luminosity
classes present in the sample, rules in or out an Outer Arm at 5--6~kpc.

\setcounter{figure}{12}
\begin{figure*}
\includegraphics[width=0.85\linewidth]{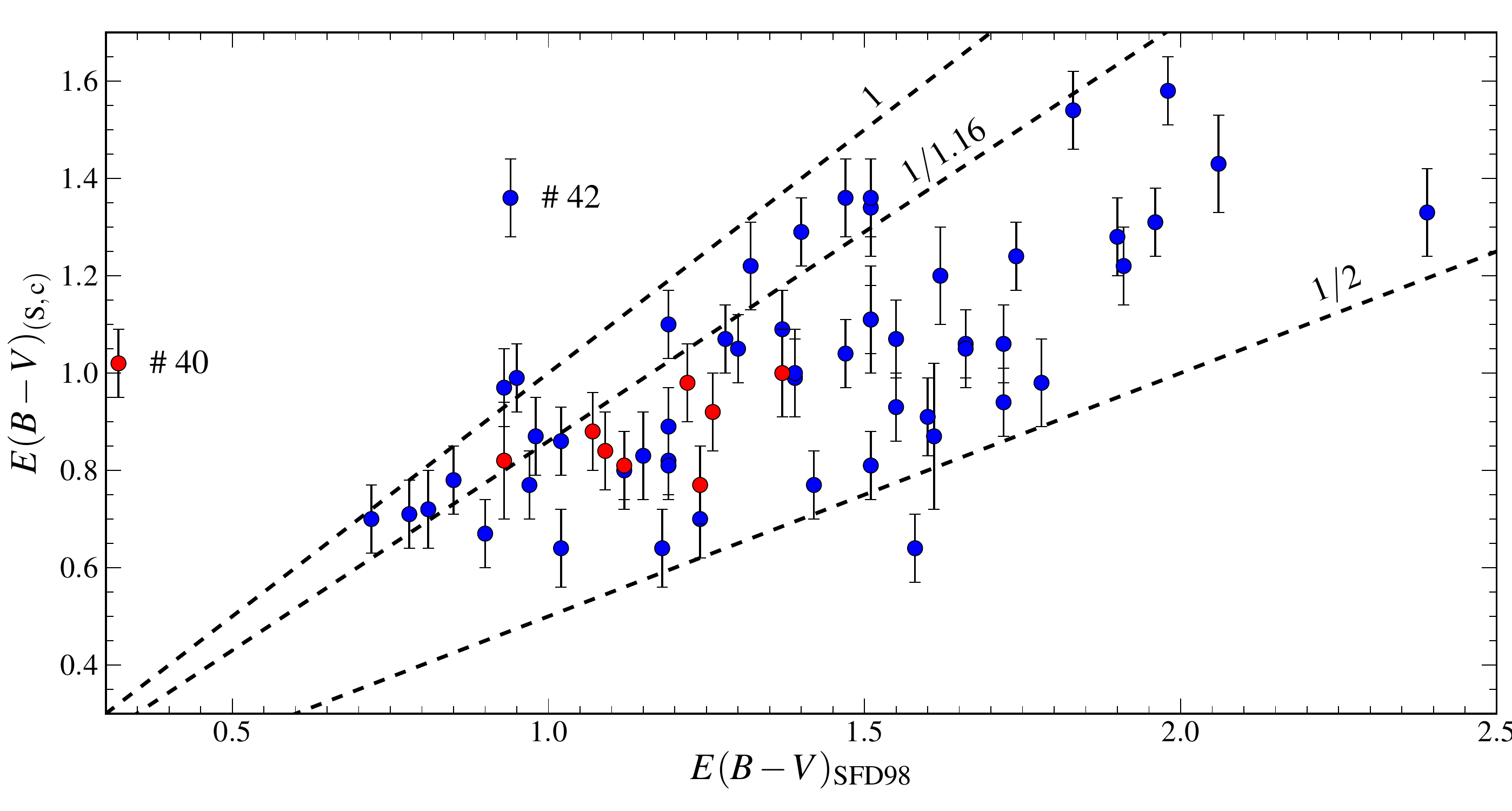}
\caption{The corrected colour excesses for the CBe star sample
  is plotted against the corresponding integrated Galactic colour
  excesses from 
  \citetalias{SFD98}, for each sightline.  Red symbols are used for the
  stars with a measured distance larger than 8~kpc, well beyond the
  expected location of the Outer Arm. The three dashed black lines are
  linear relations of the form $E(B-V)_{\rm{(S,\, c)}} = m E(B-V)_{\rm{SFD98}}$,
  where $m = 1,\, 1/1.16,\, 1/2$ respectively.}
\label{asymptotic_reddenings}
\end{figure*}
\subsection{Comparison with total Galactic colour excesses from SFD98}
\label{chap5.3}

CBe stars are massive, intrinsically-luminous stars capable of being
seen to very large distances.  Among the current sample, there are several
examples of CBe stars at distances large enough to indicate
they lie beyond even the putative Outer Arm.  The reddening of such
objects ought to closely match the integrated Galactic value  since
little Galactic dust should lie beyond them in the far outer disc.
Less distant CBe stars should in general exhibit reddenings below the
total for the relevant sightline.

The most widely used source of integrated reddenings is the work of 
\citetalias{SFD98}.  We have plotted the measured colour-excess, 
$E(B-V)_{\rm{(S,\,c)}}$, for each CBe star against the colour excess, 
$E(B-V)_{\rm{SFD98}}$, provided by SFD98
(Fig.~\ref{asymptotic_reddenings}: both values are listed in 
Table~\ref{reddenings_t}).  The former is a spot value pertaining to a 
single line of sight, while the latter applies to a spatial resolution 
element about 6 arcmins across.  Hence it should be kept in mind that
small variations in the ISM will add a random noise element to the 
comparisons we make.   We continue to exclude the four stars that do
not have a measured $EW(\Halpha)$ in Table~\ref{reddenings_t}.    A
general property of the diagram is that for all but two objects, 
$E(B-V)_{\rm{SFD98}} > E(B-V)_{\rm{(S,\,c)}}$, to within the
errors. This accords with expectation.  

The stars shown in red are the most distant, at more than 8~kpc away,
whose reddenings should most nearly match the total Galactic value
(these are specifically, objects \#~11, 26, 35, 39, 40, 50, 54, 55,
and 60).  All but one of these stars (\#~40) are early-B dwarfs.  
However, apart from \#~40, we find that the
measured colour excesses for these objects are distinctly less than
those from the \citetalias{SFD98} reddening map.  The discrepancy is
of order 0.2-0.3 magnitudes.  This undershoot is broadly in keeping
with the result of \cite{Chen99} that SFD98 typically overestimate the 
reddening by a factor of 1.16.  To illustrate this, a second reference line is
drawn in figure~\ref{asymptotic_reddenings} with this correction applied.

For object \#~40 the situation is quite different, since the datum from
\citetalias{SFD98} indicates a very much lower total dust column than
we obtain. We notice in the SFD98 temperature map (that is much less
well-resolved spatially than the emissivity map) a large hot 
spot roughly corresponding to the upper part of the galactic chimney 
linked to W4 \citep[cf.][]{Normandeau00, Terebey03}: it seems plausible therefore that the cause
of the problem is the adoption of too high a dust temperature for this 
particular sightline predicting too low a dust column.  A similar but 
not so extreme discrepancy arises in the case of object \#~42.

A further group of stars can be picked out in 
Fig.~\ref{asymptotic_reddenings}, whose colour excesses fall above
the lower reference line but remain compatible with or below the SFD98
equality line.  They are objects \#~17, 19, 22, 32, 41, 44, 49, 51,
52, 61, 64. As their estimated distances either exceed $\sim 6$~kpc,
or their sightlines are at latitudes higher than $b = 2^{\circ}$ it is
conceivable these objects also lie beyond most/all of the dust column.  
Alternatively if the scaling down of the SFD98 reddening by a factor
of 1.16 is consistently the better guide to the total Galactic value, it might be
concluded our reddenings for these stars are too high, perhaps through
under-correction for the circumstellar disc contribution, and their 
estimated distances too low.     
 
A clear feature of the sample as a whole is that their measured 
reddenings are a significant fraction of the sightline total,
ranging from about half the SFD98 value up to rough equality with it.
These large fractions of the total dust columns are to be expected given 
the long sightlines to these intrinsically bright objects.  

\section{Conclusions}
\label{chap6}
In this study, we investigated a 100~deg$^{2}$ portion of the Galactic Plane, 
between $120^{\circ} \leq \ell \leq 140^{\circ}$ and 
$-1^{\circ} \leq b \leq 4^{\circ}$, that includes a part of the
Perseus Arm, $\sim$2~kpc away and of the less well-established Outer 
or Cygnus Arm, 5--6~kpc distant.

We studied a group of 67 candidate classical Be stars that we selected
among 230 that in turn were selected from follow-up of candidate
emission line stars. We determined their spectral types with an estimated
accuracy of $\pm 1$ sub-type and measured colour excesses via SED fitting
in the blue (3800~--~5000~\AA), and made appropriate correction
for the contribution to the colour excess for circumstellar emission. 
Distances were determined via spectroscopic parallaxes, 
after luminosity classes had been assessed using MS fits to A/F stars of 
similar reddening selected via colour-cuts from IPHAS photometry, 
in the vicinity of each CBe star. Our main findings are:
\begin{itemize}

\item IPHAS offers very effective, easy selection of moderately
  reddened ($E(B-V)\sim 1$) classical Be candidates: their identity
  has been confirmed by a combination of low resolution optical
  spectroscopy and infrared photometry.

\item Our magnitude limited sample ($r \leq 16$) includes 10--15 
  stars in the outer disc at Galactocentric
  radii where the stellar density gradient is likely to be steepening 
  ($R_{\rm{G}} \geq 13$~kpc, or heliocentric distances greater than
  7~kpc).  These objects exhibit reddenings comparable with those 
  obtained from the map of \citet{SFD98}, serving to emphasise how
  far out in the Galactic disc they are.  

\item  The errors on the distance estimates remain
  too large to obtain a decisive statistical test of models for the
  spatial distribution of the CBe stars.  The major, presently irreducible, contribution to the error 
  budget is the spread of absolute magnitude associated with a given spectral type. 

\end{itemize}

These first results will be investigated in more depth, using a much 
larger sample of $\sim 200$ classical Be star candidates observed with
FAST, aided by extinction-distance curves built from IPHAS photometry 
\citep[using a new Bayesian implementation of MEAD,][]{Sale12}. This 
approach has the potential to provide a better grip on both luminosity class 
and distance even in the absence of precise absorption-line diagnostics.
The longer term prospect is that astrometry returned by the Gaia
  mission, due to launch in 2013, will greatly improve the distances
  estimated for samples of objects like the one presented in this
  study.  From predicted end-of-mission performance 
data \citep{deBruijne12}, it appears we can look forward to $5-10$\% parallax errors for our 
  objects perhaps within the decade, as compared with up to 20\%
  presently.  Especially when accompanied by carefully measured 
extinctions, that are essential for clarifying intrinsic absolute magnitudes,
further enlargement of the sample of well-characterised
fainter CBe stars will provide an avenue to better test our knowledge
both of Galactic structure and of massive-star evolution. 

\section*{Acknowledgments}
 This paper makes use of data obtained as part of 
IPHAS carried out at the Isaac Newton Telescope (INT). The INT is
operated on the island of La Palma by the Isaac Newton Group in the 
Observatorio del Roque de los Muchachos 
of the Instituto de Astrofisica de Canarias. All IPHAS data are processed 
by the Cambridge Astronomical Survey Unit, at the Institute of Astronomy
in Cambridge. We also acknowledge the use of data obtained at the INT
and the Nordic Optical Telescope as part of a CCI International Time 
Programme.  The low-resolution spectra were
obtained at the FLWO-1.5m with FAST, which is operated by
Harvard-Smithsonian Centre for Astrophysics. 
In particular, we would like to thank Perry Berlind and Mike Calkins for 
their role in obtaining most of the FLWO-1.5m/FAST data. 
RR acknowledges the University of Hertfordshire for the 
studentship support. The work of JF is supported by the Spanish
Plan Nacional de I+D+i and FEDER under contract AYA2010-18352, 
and partially supported by the Generalitat Valenciana project 
of excellence PROMETEO/2009/064.
DS acknowledges an STFC Advanced Fellowship.
Support for SES is provided by the Ministry for the Economy,
Development, and Tourism's Programa Iniciativa Cient\'{i}fica Milenio
through grant P07-021-F, awarded to The Milky Way Millennium Nucleus.

\bibliographystyle{mn2e}

\end{document}